\documentclass{aastex631}

\usepackage{colortbl}
\usepackage{printlen}
\usepackage{hyperref}

\def\s{\phantom}

\def\amin{\ifmmode^{\prime}\else$^{\prime}$\fi}
\def\asec{\ifmmode^{\prime\prime}\else$^{\prime\prime}$\fi}

\def\simgt{\lower.5ex\hbox{$\; \buildrel > \over \sim \;$}}
\def\simlt{\lower.5ex\hbox{$\; \buildrel < \over \sim \;$}}

\newcommand\rosat{{\it ROSAT\/}}

\newcommand\xmm{{\it XMM-Newton}}

\newcommand\swift{{\it Swift\/}}
\newcommand\nustar{{\it NuSTAR}}
\newcommand\nicer{{\it NICER}}

\newcommand{\ms}{$M_{\odot}$}

\newcommand{\fluxcgs}{ergs~s$^{-1}$~cm$^{-2}$}
\newcommand{\lumcgs}{ergs~s$^{-1}$}
 
\newcommand{\mdot}{$\dot{m}$} 

\begin{document}

\title{NuSTAR broadband X-ray observation of EF Eri following its reawakening into a high accretion state}

\author[0009-0004-9156-7893]{Luke W. Filor}
\affiliation{Columbia Astrophysics Laboratory, Columbia University, New York, NY 10027, USA}

\author[0000-0002-9709-5389]{Kaya Mori}
\affiliation{Columbia Astrophysics Laboratory, Columbia University, New York, NY 10027, USA}

\author[0000-0002-6653-4975]{Gabriel Bridges}
\affiliation{Columbia Astrophysics Laboratory, Columbia University, New York, NY 10027, USA}

\author[0000-0002-3681-145X]{Charles J. Hailey}
\affiliation{Columbia Astrophysics Laboratory, Columbia University, New York, NY 10027, USA}

\author[0000-0002-7004-9956]{David A. H. Buckley}
\affiliation{South African Astronomical Observatory, P.O Box 9, Observatory, 7935 Cape Town, South Africa}
\affiliation{Department of Astronomy, University of Cape Town, Private Bag X3, Rondebosch 7701, South Africa}
\affiliation{Department of Physics, University of the Free State, PO Box 339, Bloemfontein 9300, South Africa}

\author[0000-0001-8722-9710]{Gavin Ramsay}
\affiliation{Armagh Observatory and Planetarium, College Hill, Armagh, BT61 9DG, UK}

\author[0000-0003-3441-9355]{Axel D. Schwope}
\affiliation{Leibniz-Institut für Astrophysik Postdam (AIP), An der Sternwarte 16, 14482 Potsdam, Germany}

\author[0000-0003-3733-7267]{Valery F. Suleimanov}
\affiliation{Institut für Astronomie und Astrophysik, Universität Tübingen, Sand 1, 72076 Tübingen, Germany}

\author[0000-0002-4013-5650]{Michael T. Wolff}
\affiliation{Space Science Division, U.S. Naval Research Laboratory, Washington, DC 20375, USA}

\author{Kent S. Wood}
\affiliation{Praxis Inc., Alexandria, VA 22303, USA}
\affiliation{Resident at Naval Research Laboratory, Washington, DC 20375, USA}

\correspondingauthor{Luke W. Filor, Kaya Mori, Gabriel Bridges}  \email{lwf2113@columbia.edu, kaya@astro.columbia.edu, glb2139@columbia.edu}

\begin{abstract}

We present the first \nustar\ X-ray observation of EF Eri, a well-known polar system.  The \nustar\ observation was conducted in conjunction with \nicer\, shortly after EF Eri entered a high accretion state following an unprecedented period of low activity lasting 26 years since 1997. \nustar\ detected hard X-ray emission up to 50 keV with an X-ray flux of $1.2\times10^{-10}$~\fluxcgs\ (3--50 keV). Folded X-ray lightcurves exhibit a single peak with $\sim65$\% spin modulation throughout the 3--50 keV band. We found no evidence of QPO signals at $\nu = 0.1\rm{-}100$ Hz with an upper limit on the QPO amplitude below 5\% (90\% CL) at $\nu \sim 0.5$ Hz where the optical QPO was previously detected. Our 1-D accretion column model, called {\tt MCVSPEC},  was fitted to the \nustar\ spectral data, yielding an accurate WD mass measurement of $M = (0.55\rm{-}0.63) M_\odot$. \texttt{MCVSPEC} accounts for radiative cooling by thermal bremsstrahlung and cyclotron emission, X-ray reflection off the WD surface, and a previously constrained range of the accretion column area. The derived WD mass range is in excellent agreement with the previous measurement of $M = (0.55\rm{-}0.65) M_\odot$ in the optical band.  This demonstrates a combination of broadband X-ray spectral analysis and the {\tt MCVSPEC} model that can be employed in our ongoing \nustar\ observation campaign of other polars to determine their WD masses accurately.

\end{abstract}


\section{Introduction} \label{sec_intro}
Polars are magnetic cataclysmic variables (mCVs) consisting of a highly magnetized white dwarf (WD) and a late main-sequence secondary star. Due to the strong WD magnetic field strengths observed in the range of $B = 7\rm{-}240$ MG, their spin and orbital periods are synchronized and accretion disk formation is prohibited \citep{cropper1990polars}. In-falling gas from the companion star is channeled along magnetic field lines toward the WD polar caps. The free-falling gas is heated to high plasma temperatures ($kT_{\rm s} \simgt 10$ keV) as it becomes supersonic at a stand-off shock. Below the shock, there exists a column of material cooling via thermal bremsstrahlung and cyclotron radiation toward the WD surface, exhibiting a range of temperatures and densities \citep{imamura1987x}.  As a result, polars emit copious thermal X-rays from accretion columns, while their optical emission is usually dominated by cyclotron radiation and WD surface emission \citep{lamb1979x, Mukai2017}. Due to the absence of accretion disks, polars are highly variable both in the optical and X-ray bands as they alternate between low and high accretion states.  

EF Eri is the fourth polar discovered after AM Her, VV Pup, and AN UMa, and it is one of the best-studied objects in the optical, UV, and X-ray bands. EF Eri's WD has a spin period of $P=81$ minutes and a magnetic field strength of $B\sim13$ MG at the accretion region \citep{patterson1981amazing, campbell2008cyclotron}.  EF Eri is well known for its extremely prolonged low accretion state lasting for 26 years since 1997. Before the low accretion state, EF Eri had been extensively observed in the X-ray band dating back to the first X-ray detection by {\it Ariel V} \citep{cooke1978ariel}. Subsequent X-ray observations by {\it Einstein, EXOSAT, Ginga} and \rosat\ established its strong thermal X-ray emission modulated with the WD spin period \citep{patterson1981amazing, watson_exosat, done_ginga, Beardmore_ginga_qpos, beuermann1991short}.  

EF Eri is one of the few polars that exhibited 1--3-sec optical QPOs and the only polar to exhibit X-ray QPOs at much lower frequencies than the optical QPOs, reported by {\it Einstein} and {\it EXOSAT} observations. These optical QPO signals have been attributed to the instability of accretion flow (e.g., \citealt{larsson1987discovery}).  The X-ray QPOs found in the 1979 \textit{Einstein} observation had a period of 6 minutes in the 0.1--4.0 keV band \citep{patterson1981amazing}.  These 6-minute QPOs were also reported in optical photometry, thought to arise from unstable mass accretion onto the WD \citep{williams1980light}. 
An \textit{EXOSAT} observation from 1983 found 4-minute (2--6 keV) X-ray QPOs with a modulation of 10\% \citep{watson_exosat}. Optical observations performed simultaneously with \textit{EXOSAT} by the CTIO telescope revealed a peak coincident with the X-ray quasi-period of 4 minutes. Conversely, \textit{Ginga} and \rosat\ observations performed between 1988-1990 found no evidence for similar X-ray QPOs \citep{done_ginga, beuermann1991short}.  The existence of X-ray and optical QPOs on the 4--6 minute scale in addition to rapid optical QPOs between 1--3 seconds remains puzzling. 

In 1997, EF Eri entered a prolonged low state as characterized in \citep{imamura2000coordinated}, a few years before the arrival of the more sensitive \textit{XMM-Newton} and \textit{Chandra} X-ray telescopes. Optical photometric and spectroscopic observations in early 1997 showed EF Eri to be four magnitudes fainter than usual ($\sim$ 18 mag), with its optical modulation being reduced to $\sim 0.2$ mag \citep{wheatley1998ef}.  Unlike the {\it Ginga} (1.8--18 keV) X-ray detection during the high state, \textit{RXTE} failed to detect EF Eri in the X-ray band.  Multiple observations performed by \xmm\ found that EF Eri's X-ray flux had diminished by three orders of magnitude to $F_X \sim 1.5-7 \times {10}^{-14}$ \fluxcgs\ \citep{schwope2007xmm}.  This period of low accretion was only interrupted by two brief instances of increased accretion in 2006 and 2008 \citep{mukai2008end}.  This prolonged, consistent low state lasted for 26 years, leading to EF Eri being referenced as a textbook example of low-accretion rate polars (LARPs). 

\begin{figure}[bh] 
\begin{center}
\includegraphics[width=0.7\textwidth]{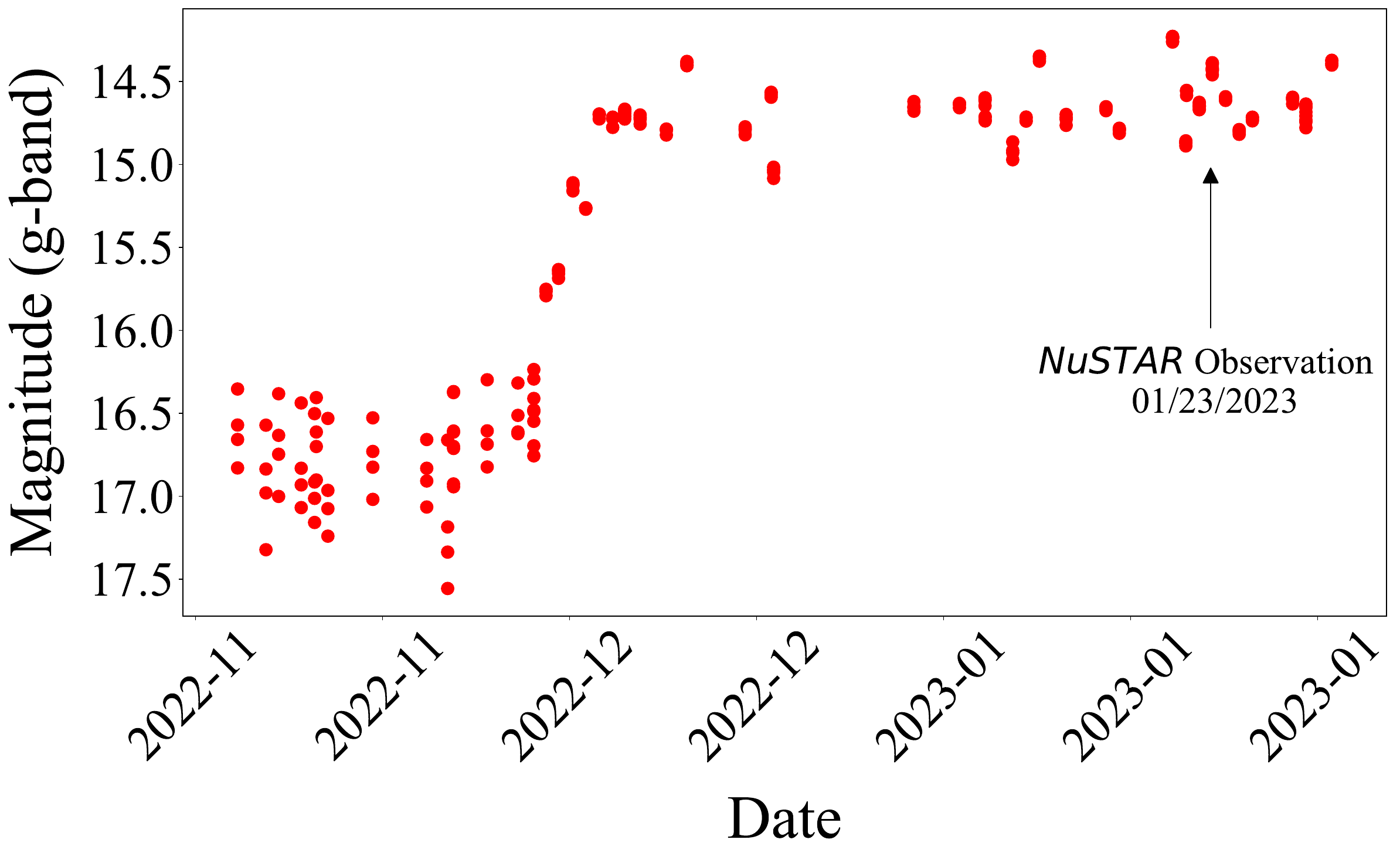}
\caption{EF Eri's g-band lightcurve from 11/2022 to 02/2023.  EF Eri displayed an optical brightening of several magnitudes, corresponding to increased accretion of material onto the WD surface.  This lightcurve was created using the ASASSN Sky Patrol Photometry Database \citep{kochanek2017all, shappee2014man}.}
\label{fig_ASASSN}
\end{center}
\end{figure}

In late 2022, the historically low state concluded abruptly as EF Eri awoke into a high state as its optical brightness increased by 3 magnitudes over only 2 weeks, once again becoming one of the brightest polars in the sky at 14.5--15 mag (g-band).
The optical brightening was monitored by AAVSO and ASAS-SN establishing a new high accretion state (Figure \ref{fig_ASASSN}) \citep{kochanek2017all,shappee2014man}. After confirming EF Eri to be in a high state through a \swift-XRT observation, we performed joint \nustar\ and \nicer\ ToO observations of EF Eri. 

EF Eri is part of the ongoing \nustar\ observation campaign (PI: K. Mori) by monitoring $\sim 40$ polars and triggering \nustar\ ToO observations when a source enters a high state.  The primary goal of this program is to determine the WD masses of polars using broadband X-ray spectral data. Our paper presents the first \nustar\ observation of EF Eri in a high state, which is also the first observation under the \nustar\ ToO campaign. Utilizing the high-quality broadband X-ray timing and spectral data obtained by \nustar\, we search for X-ray QPOs, characterize broadband X-ray spectra and their phase variation, and determine the WD mass accurately.    

Our paper is structured as follows. 
In \S\ref{sec_observations}, we describe the \nustar\ and \nicer\ X-ray observations and data reduction methods.  In \S\ref{sec_timing}, we present X-ray lightcurve analysis and search for X-ray QPOs such as those found in the \textit{Einstein} and \textit{EXOSAT} observations.  In \S\ref{sec_spectral}, we fit phenomenological models to the \nustar\ phase averaged and phase resolved spectral data. In \S\ref{sec_mass} we determine the WD mass using our accretion column model ({\tt MCVSPEC}) specifically developed for polar X-ray emission.  In \S\ref{sec_disc}, we discuss the implications of the first hard X-ray observation of EF Eri's high state and the newly derived WD mass.   In \S\ref{sec_summ}, we summarize our findings and plan for observing other polars through the \nustar\ ToO program.
\begin{deluxetable*}{lcccc}[ht!] \label{tab_xrayobs}

\tablecaption{Summary of Major EF Eri X-ray Observations}
\tablecolumns{5}
\tablehead{
\colhead{Observatory}   
& 
\colhead{Energy Band}
& 
\colhead{Date}
&
\colhead{EF Eri State}
&
\colhead{Reference}
}
\startdata  
Ariel V  & $2-18$ keV     & 1978    & H & \citet{cooke1978ariel} \\
Einstein & $1.2 - 10$ keV & 07/1979 & H & \citet{patterson1981amazing} \\
EXOSAT & $0.04 - 2$ keV & 11/1983 & H & \citet{watson_exosat} \\
Ginga & $1.8 - 18$ keV & 12/1988 & H & \citet{done_ginga}, \citet{Beardmore_ginga_qpos} \\
ROSAT & $0.06-2.4$ keV & 07/1990 & H & \citet{beuermann1991short} \\
RXTE* & $2.0-60$ keV & 08/1996 & L & \citet{imamura2000coordinated} \\
XMM & $0.1-10$ keV  & 08/2002 - 02/2003 & L & \citet{schwope2007xmm} \\
NICER & $0.2-12.0$ keV & 01/20/2023 - 01/30/2023 & H & {This Paper}\\
{NuSTAR} & {$3-79$ keV} & {01/23/2023} & {H} & {This Paper}\\
\enddata
\textbf{$^*$} No X-ray detection.\\
\end{deluxetable*}

\section{Observations} \label{sec_observations}

Upon the detection of an optical brightening first reported by AAVSO (see Figure \ref{fig_ASASSN} for a g-band optical lightcurve obtained by ASAS-SN), a 1--ks \swift\ ToO observation was triggered on 1/16/2023 to confirm the optical brightening corresponded with an X-ray brightening.  The \swift\ observation (\textbf{ObsID: 00031180015}) recorded an absorbed flux of $1.1\times{10}^{-10}$~\fluxcgs\ (0.3--10 keV), three orders of magnitude higher than the X-ray fluxes measured by \textit{XMM-Newton} in 2002-2003, confirming the onset of a new high accretion state. Subsequently, a 22--ks \nustar\ ToO observation was conducted on 1/23/2023 (\textbf{ObsID: 90901303002}).  Simultaneously, several \textit{NICER} observations were conducted in the days surrounding the \nustar\ ToO.  One 4.6--ks \textit{NICER} observation (\textbf{ObsID: 5141010102}) taken on 1/23/2023 was fully concurrent with the \nustar\ observation.

We processed \nustar\ data using  \textbf{nupipeline} in the \nustar\ Data Analysis Software (\texttt{NuSTARDAS} version 2.1.2). We extracted source lightcurve and spectral data from a $r = 60$\asec\ circular region around the source position. Background data were extracted from a source-free region on the same detector chip where the source is located. The \nustar\ observation detected the source above background level up to 50 keV and collected a total of $\sim 55,000$ counts in 3--50 keV and $\sim12,000$ counts in 10--50 keV after combining \texttt{FPMA} and \texttt{FPMB} data.  

We processed the \nicer\ data using the \textbf{nicerl2} and \textbf{nicerl3-spect} tasks in the \nicer\ Data Analysis Software (\texttt{NICERDAS} version 13).  The \textbf{SCORPEON} background model (version 23) was used when fitting the spectral data.  An energy range of $0.3 - 12$ keV was considered, as the region of the NICER spectrum below 0.3 keV introduced a degeneracy between a normalization parameter in the \textbf{SCORPEON} background model and the galactic hydrogen column density.

\section{Timing Analysis} \label{sec_timing}

We utilize the X-ray timing analysis software {\tt Stingray} and {\tt HENDRICS} \citep{matteo_bachetti_2022_6394742, hendrics} for generating lightcurves and power density spectra (PDS) of EF Eri. We apply a barycentric correction to photon arrival times using the {\tt HEASOFT} tool {\tt barycorr v2.17}\citep{heasoft} (using DE405 solar system ephemeris) before extracting events from a $60"$ circular region. We observe an average count rate of $1.22$ ($1.17$) cnts/s in FPMA (FPMB); we estimate $98.7\%$ of these photons come from EF Eri for both focal plane modules. These event lists are calibrated using the {\tt HENDRICS} command-line utilities, and soft (3--10 keV) and hard (10--50 keV) lightcurves are extracted.
These lightcurves show significant variability ($\sim 43\%$), which corresponds well to the orbital ephemeris of EF Eri.

\subsection{Power Density Spectra} \label{sec_pds}

\begin{figure}[b]
    \centering
    \includegraphics[width=\linewidth]{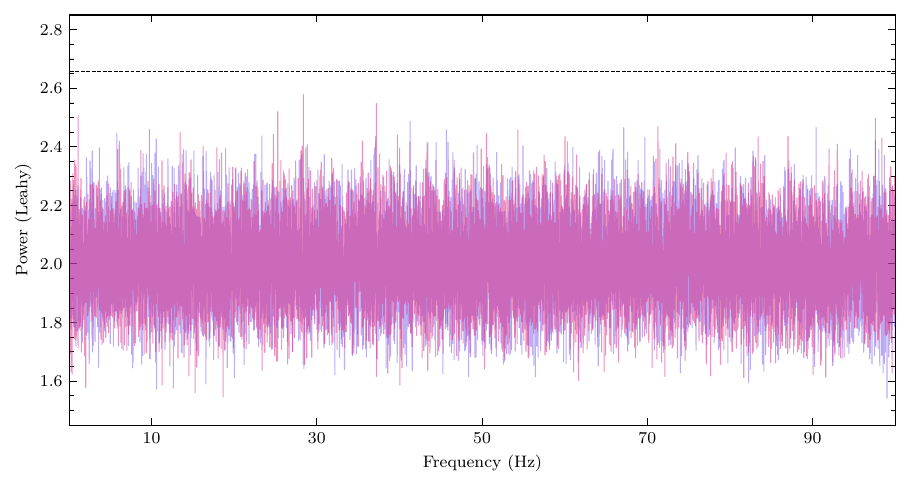}
    \caption{Deadtime-corrected power density spectra for FPMA (pink) and FPMB (purple). The PDS are averaged from 221 lightcurve segments with a length of 104 s each.  The horizontal dashed lines indicate the 3$\sigma$ detection threshold.}
    \label{fig:fig_pds}
\end{figure}

\nustar\ suffers from a long ($\sim 2.5$ ms), and non-constant, deadtime, which can cause distortions to the PDS \citep{bachetti2015}. We follow the methods of \cite{bachetti2018} to produce Leahy-normalized deadtime-corrected PDSs for each \nustar\ focal plane module and energy band. We construct lightcurves with a binning time of $\sim3.9$ ms (chosen to permit a NyQuist frequency of $>100$Hz and a bin time that is a power of 2 for optimized use of the FFT algorithm). Each PDS is the product of averaging 221 individual PDSs made from lightcurve segments of length 104 s. Inspection of the individual PDSs made from each segment reveals that, despite the variability in the lightcurve, the PDS spectral shape is constant throughout the observation. 
 
The averaged PDS lacks distinct features and appears to be entirely white noise. Nevertheless, we search for a potential QPO signal in the 0.1--100 Hz range \citep{VanBoxSom2017}. White noise in Leahy normalized PDSs are $\chi^2_{2M}/M$ distributed \citep{leahy1983}. The most significant power in the 0.1--100 Hz range does not exceed $0.9\sigma$ significance. 

To derive an upper limit of the QPO amplitude, we performed PDS simulation, assuming a QPO signal at $\nu = \nu_0$ in the Lorentizan functional form of $P_{\rm QPO} = \frac{A \Delta}{\pi(\Delta^2+(\nu-\nu_0)^2}$ where $A$ represents the amplitude and $\Delta$ is the line width. We added this QPO signal to the white-noise level (Figure \ref{fig:fig_pds}) within the frequency range of our interest ($\nu \simgt 0.1$ Hz), where the optical QPOs have been detected and X-ray QPOs have been predicted \citep{VanBoxSom2017} (discussed later in \S\ref{subsec_timingdisc}). For a given number of segments $M$, we generate a random number from the  $\chi^2_{2M}/M$ distribution and multiply it to $1+ R P_{\rm QPO}$ where $R$ is the source count rate ($\sim1$ cts\,s$^{-1}$). We repeated this procedure $10^4$ times and recorded whether the QPO signal exceeded the 3-$\sigma$ threshold (dashed lines in Figure \ref{fig:fig_pds}), in which case we tagged a detection. We fixed $\nu_0 = 0.5$ or $\nu_0 = 10$ Hz, which correspond to the observed optical QPO centroid for EF Eri \citep{larsson1987discovery} and the X-ray QPO frequency range predicted by a recent MHD simulation work \citep{VanBoxSom2017}, respectively. For the so-called quality factor defined by $Q \equiv \nu_0/\Delta$, we adopted $Q = 1\rm{-}3$ as a reasonable range from the optical QPOs observed from five polars \citep{Beardmore1997}.  
From the PDS simulations for $\Delta T = 201$ [s] segment length, we determined a 
90\% limit amplitude to be $A < 7$\% and $< 80$\% for $\nu = 0.5$ and 10 Hz, respectively.  
The QPO amplitude upper limit is not so well constrained at $\nu_0 = 10$ Hz as the sampling rate of $1/R \sim 1$ Hz is much smaller than 10 Hz. 

Searching for high-frequency QPOs at $\nu_0 \gg 1$ Hz will require significantly higher count rates, for example, from the \nicer\ data.  The search for QPOs in the \nicer\ data will be presented in a subsequent publication.

\subsection{Orbital Period Search} \label{orb_period}

\begin{figure} 
    \centering
    \includegraphics[width=\textwidth]{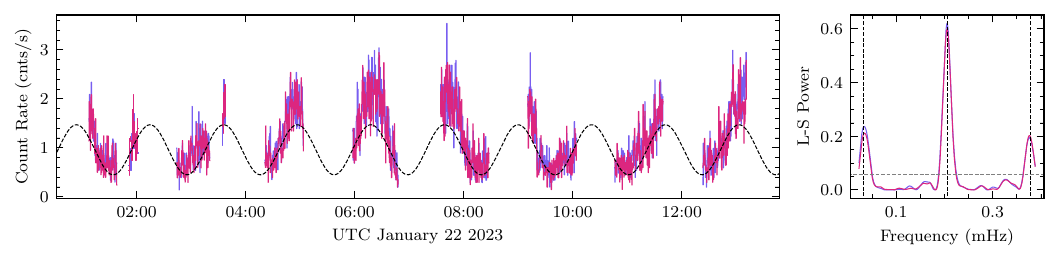}
    \caption{Left: $\nustar$ FPMA (pink) and FPMB (purple) 3--50 KeV light-curves. A sine wave is overlaid with the ephemeris taken from \cite{schwope2010x}.  Right: Lomb-Scargle periodograms for FPMA and FPMB for 0.23--0.388 mHz. Dashed vertical lines indicate the 81.02 min spin period of EF Eri and the adjacent aliases with the NuSTAR 96 min orbital period. The Horizontal dashed line indicates the 5$\sigma$ detection threshold.}
    \label{fig_period}
\end{figure}

EF Eri has an orbital period of $P_{\mathrm{orb}} = 81.02$ minutes, comparable to the \nustar\ orbital period of $P_{\mathrm{NuSTAR}} \sim 96$ minutes. A powerful way to search for low-frequency signals in data with gaps (due primarily to satellite occultation in this case) is the Lomb-Scargle Periodogram, illustrated in Figure \ref{fig_period} \citep{Lomb1976, Scargle1982}. We extract one lightcurve across the entire (3--50 keV) energy band from each focal plane module with a binning time of $20$ s (chosen to ensure $\sim 10$ counts per bin) and construct a Lomb-Scargle Periodogram using the {\tt nifty-ls} software package's {\tt Astropy} interface \citep{niftyls, astropy:2022}. We search between 0.023--0.388 mHz with an oversampling factor of 32 and find a very strong peak centered at $P_{\mathrm{orb}} = 81$ minutes. Two weaker, but still significant, peaks are detected at the alias frequencies $f_{\mathrm{orb}}\pm f_{\mathrm{NuSTAR}}$. The central peak is detected at 30$\sigma$ significance and has a FWHM of $9.1\%$.

\subsection{Accretion Geometry} \label{Folded LC}

\begin{figure} 
    \centering
    \includegraphics[width=\textwidth]{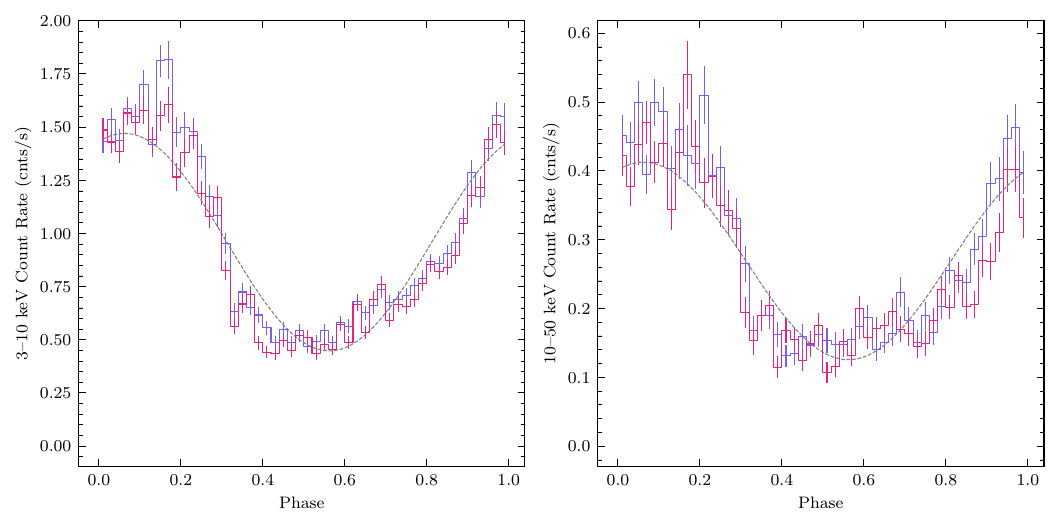}
    \caption{\nustar\ FPMA (pink) and FPMB (purple) 3--10 and 10--50 keV folded light-curves. Lightcurves are folded on the EF Eri orbital period from \cite{bailey1982} and phased from the ephemeris in \cite{schwope2010x}. The folded lightcurves are overlayed with the best-fit sinusoidal function in gray.} 
    \label{fig_foldedlc}
\end{figure}

We fold the 3--10 and 10--50 keV background-subtracted lightcurves on the orbital period of EF Eri. We use the ephemeris of \cite{schwope2010x} to fix our profile to the inferior conjunction of the donor star. The pulse profile shape is energy-independent and features a sharp drop at phase $\sim$ 0.3 and a slow rise between phase 0.7 and 1. The asymmetry in this profile can not be explained by changes in the visibility of the accretion region throughout the orbit of EF Eri. These lightcurves can be seen in Figure \ref{fig_foldedlc} along with a best-fit sinusoid to illustrate the asymmetry. For further analysis, we consider the full 3--50 keV folded lightcurve.  
We compute the pulsed fraction as PF = $\frac{R_{\text{max}}-R_{\text{min}}}{R_{\text{max}}+R_{\text{min}}}$ (where $R_{\text{max}}$ ($R_{\text{min}}$) is the maximum (minimum) of the pulse profile). We take $R_{\text{max}}$ to be the average rate between phases 0.0--0.2 and $R_{\text{min}}$ to be the average rate between phases 0.4--0.6 and find a pulsed fraction of 50.4 $\pm$ 0.8\%.

The visibility of the accretion column is proportional to $\cos(\theta)$ where $\theta$ is the viewing angle and can be computed across the orbit for a given orbital inclination ($i$) and magnetic colatitude ($b$): $\cos(\theta) = \cos(i)\cos(b)-\sin(i)\sin(b)\cos(\phi)$ ($\cos(\theta)<0$ corresponds to the occultation of the column by the WD). An expected pulsed fraction can be computed based on the visibility. 
\cite{campbell2008cyclotron} measured a magnetic colatitude for EF Eri of 6\textdegree with an orbital inclination of 58\textdegree. This viewing geometry corresponds to a pulsed fraction of only 17\% and is inconsistent with the observed profile. A magnetic colatitude of 17.5\textdegree $\pm$ 0.1\textdegree~ is required to reproduce the observed pulsed fraction. 

\begin{figure}[htb] 
    \centering
    \begin{minipage}[b]{0.45\textwidth}
        \centering
        \includegraphics[width=\textwidth]{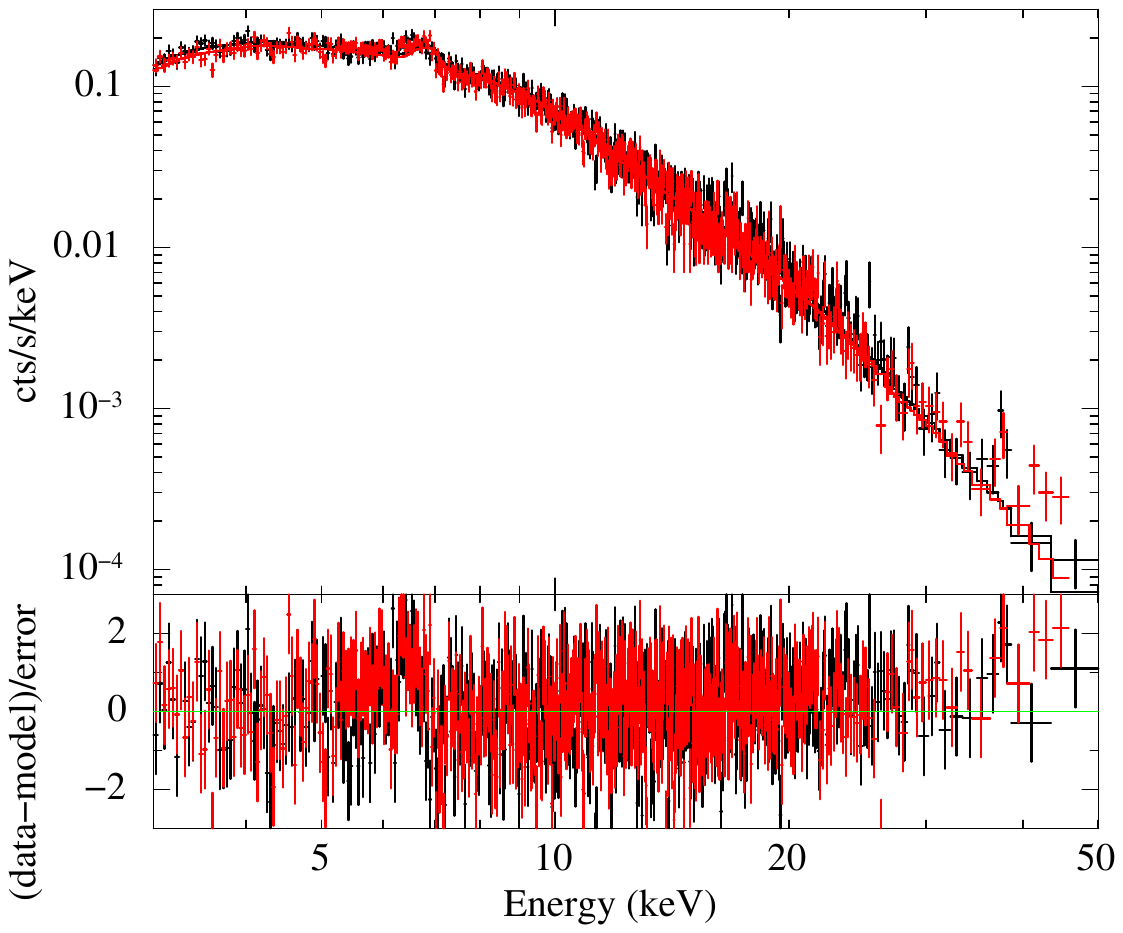}
        \label{subfig_apec}
    \end{minipage}
    \hfill
    \begin{minipage}[b]{0.45\textwidth}
        \centering
        \includegraphics[width=\textwidth]{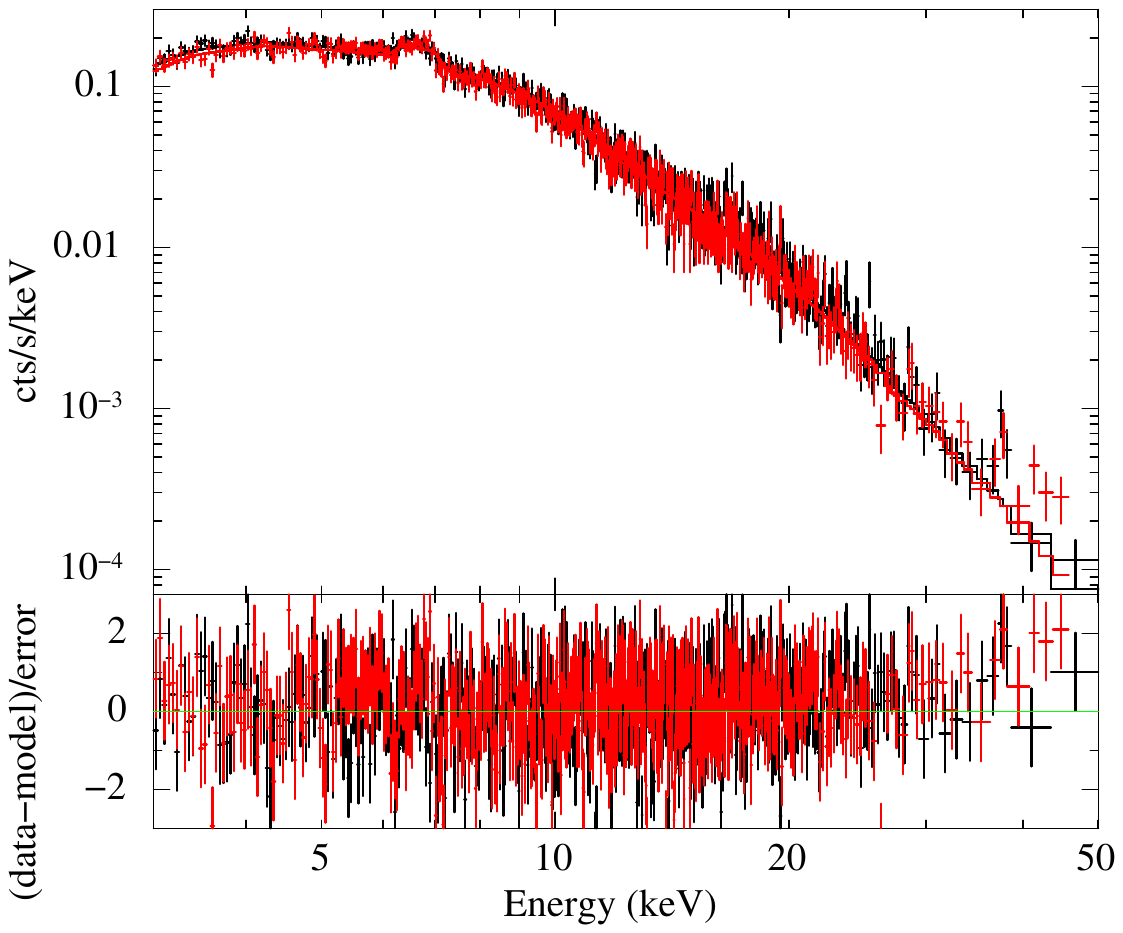}
        \label{subfig_apec+gauss}
    \end{minipage}
    \caption{Left: $\nustar$ spectra and residuals of EF Eri fit with a \texttt{tbabs*cflux*(APEC)} model, yielding a $\chi^2$ statistic of $1.02$ with $897$ degrees of freedom.  Right: \nustar\ spectra and residuals fit with a \texttt{tbabs*cflux*(APEC+gauss)} model, yielding a $\chi^2$ statistic of $0.96$ with $896$ degrees of freedom.  The $\nustar$ observation is presented in the 3--50 keV band, with FPMA in black and FPMB in red.  Parameters associated with these models are listed in Table \ref{tab_phenom}.}
    \label{fig_phenom}
\end{figure}

\section{Spectral Analysis} \label{sec_spectral}

Upon processing the \nustar\ data, we first fitted a handful of phenomenological models to characterize the overall X-ray spectral properties using XSPEC \citep{arnaud1996xspec}.  These models allowed us to estimate the X-ray luminosity as well as the average plasma temperature for the accretion column.  The current understanding of accretion column structure in polars necessitates the use of a multi-temperature model to describe the changing temperature, velocity, density, and pressure of the infalling material as it approaches the surface of the WD.  We introduce our multi-temperature model {\tt MCVSPEC} to account for the dynamic nature of the gas in the accretion column, utilizing the \texttt{reflect} model internally to account for emission reflected from the WD surface (Bridges et al. in preparation).

\begin{deluxetable*}{lcc}[!ht] \label{tab_phenom}
\tablecaption{Phenomenological Model Fits to the \nustar\ spectra}
\tablecolumns{3}
\tablehead{
\colhead{Parameter}   
& 
\colhead{{\tt APEC}}
& 
\colhead{{\tt APEC} + {\tt gauss}}
}
\startdata  
$N^{(i)}_H (10^{19} \rm{cm}^{-2})^a\;*$ & $1.0$ & $1.0$ \\
$kT_1$ (keV) & $13.3 \pm 0.4$ & $13.6_{-0.4}^{+0.3}$ \\
$Z^b (Z_\odot)$ & $0.34 \pm 0.05$ & $0.30 \pm 0.05$  \\
$E_{line}$ (keV)* & ... & $6.4$ \\
$\sigma_{line}$ (keV)* & ... & $0.01$ \\
$EW_{line}$ (eV) & ...  & $82.0 \pm 18.4$ \\
$F_X (10^{-10} \frac{\mathrm{erg}}{\mathrm{cm}^2 \mathrm{s}})^c$  & $1.15 \pm 0.01$ & $1.15\pm 0.01$ \\
$\chi^2$ (dof) & 1.02 (897) & 0.96 (896) \\
\enddata
All errors shown are $90 \%$ confidence intervals. \\
$^a$ The ISM hydrogen column density is associated with {\tt tbabs}, which is multiplied to all the models.  Due to the $3 - 79$ keV energy range of \nustar, the ISM hydrogen column density has minimal effect on the spectral fit.  $N_H$ is thus frozen to the low column density inferred from \rosat\ and {\textit EXOSAT} \citep{beuermann1991short, watson_exosat}.\\
$^b$ Abundance relative to solar. \\
$^c$ 0.1 - 100 keV unabsorbed flux of the \nustar\ data as derived from the {\tt cflux} model. \\
$^*$ The parameter is frozen. \\
\end{deluxetable*}

\subsection{Phenomenological model fits} \label{subsec_phenom}

  Each spectral model was multiplied by \texttt{tbabs} (for the ISM absorption assuming the {\tt Wilms} abundance).  The first model fit to the \nustar\ data consisted of the \texttt{APEC} model for a single-temperature plasma, in addition to the \texttt{cflux} model to estimate the unabsorbed flux of the observation.  The full model of \texttt{tbabs*cflux*(APEC)} resulted in a good fit with a $\chi^2$ statistic of $1.02$ ($897$ d.o.f.).  The fit was further improved by adding a Gaussian component to model the neutral iron K-$\alpha$ emission line at $6.4$ keV.  Figure \ref{fig_phenom} presents the best model fit and residual for an APEC model with a Gaussian line component fixed at $E = 6.4$ keV.  The best-fit plasma temperature and abundance are $kT = 13.6_{-0.4}^{+0.3}$ keV and $Z = 0.30 \pm 0.05 Z_\odot$ , respectively.  Using the \texttt{cflux} model we found an unabsorbed X-ray flux of $(1.15 \pm 0.01) \times {10}^{-10}$ \fluxcgs\ ($0.1\rm{-}100$ keV), with the full model of \texttt{tbabs*cflux*(APEC+gauss)} having a reduced $\chi^2$ value of $0.96$ (896 d.o.f.).  

\subsection{Phase-resolved spectral analysis \label{subsec_phase}} 

\begin{figure}[htbp]
    \centering
    \hfill
    \begin{minipage}[b]{0.45\textwidth}
        \centering
        \includegraphics[width=\textwidth]{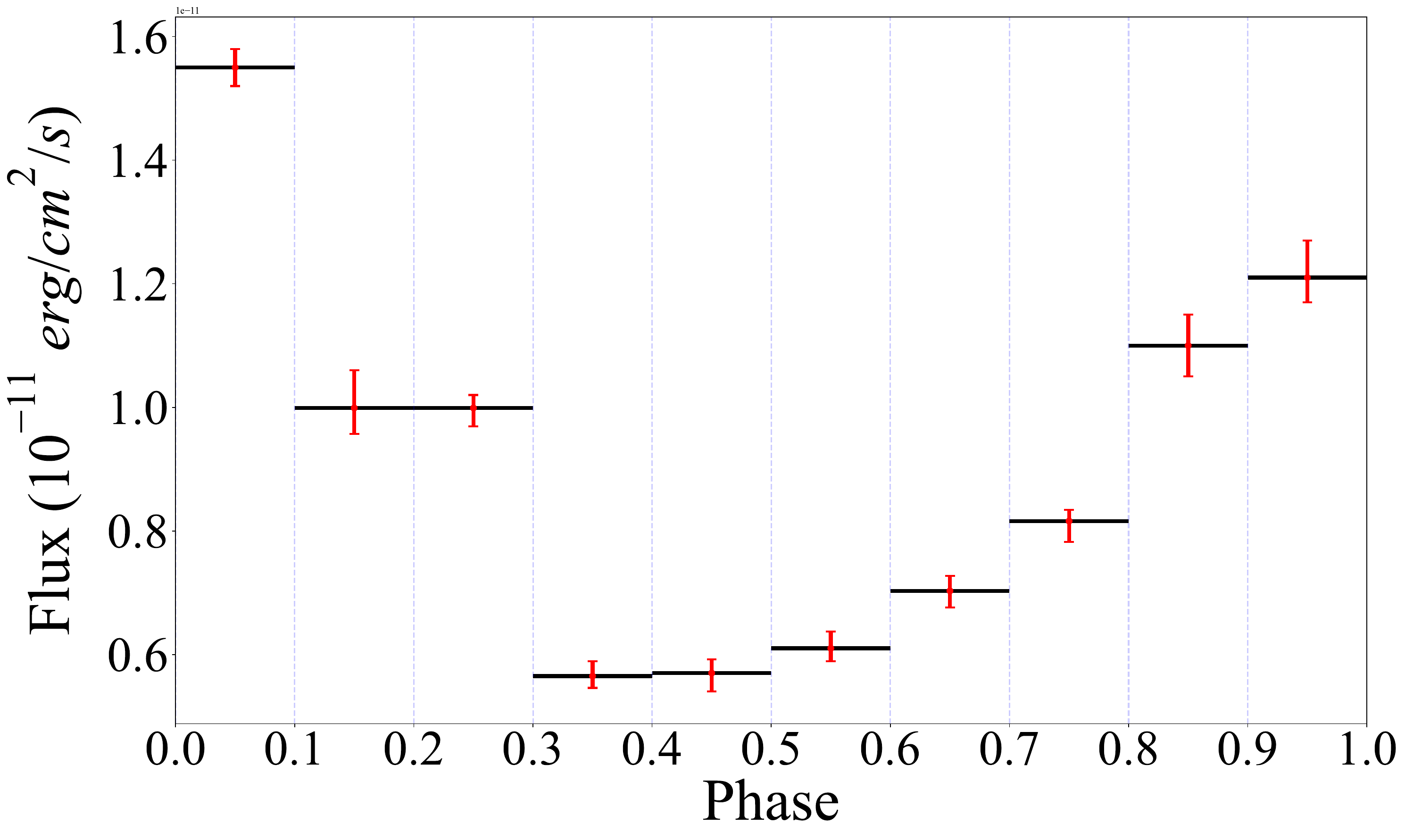}
    \end{minipage}
    \hfill
    \begin{minipage}[b]{0.45\textwidth}
        \centering
        \includegraphics[width=\textwidth]{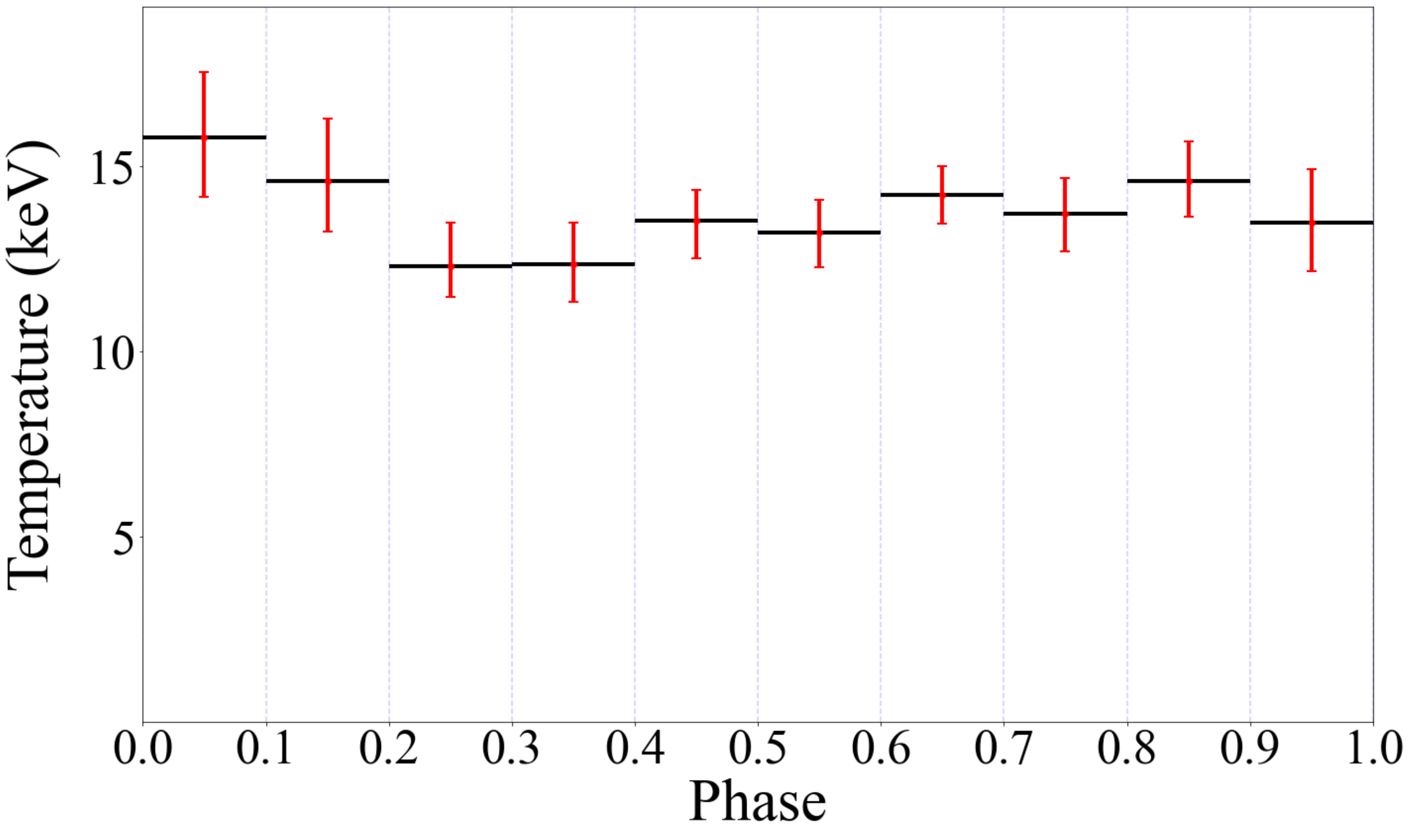}
    \end{minipage}
    \hfill
    \caption{Left: X-ray flux in the 3.0--50 keV energy range, separated across 10 phase bins spanning EF Eri's orbital period of $81$ minutes.  A significant increase in flux and photon counts is observed between phase bins $0.8 \rm{-} 0.2$.  Right: Comparison of temperature across all phases of EF Eri's orbital period.  Plasma temperature remains relatively constant between 12--16 keV.}
    \label{fig_phase}
\end{figure}

Estimates via Zeeman tomography and the spacing of cyclotron harmonics indicate an average magnetic field of $\sim 13$ MG \citep{beuermann2007zeeman, campbell2008cyclotron}.  At the same time, analysis of EF Eri suggested a local region of intense magnetic field strength ($B \gtrsim 100$ MG) localized to a small area \citep{beuermann2007zeeman, szkody2006galex, campbell2008cyclotron}.  If this local region coincides with the accretion column, we expect the X-ray spectrum to be substantially softened by increased cyclotron cooling.  
Average plasma temperature can serve as a measure of spectral hardness.  We divided the \nustar\ spectra into 10 phase bins and individually fit an \texttt{APEC} model to determine the variability of the plasma temperature. 
Figure \ref{fig_phase} illustrates the unabsorbed flux and $kT$ associated with each phase bin.  In general, the plasma $kT$ remained between 12--16 keV with little phase variation. 
Analysis of phase-resolved spectroscopy performed on EF Eri suggests that an accretion column is continuously viewable from Earth.  Therefore, we will continue our spectral analysis using a phase-averaged spectrum, as all phases clearly illustrate bremsstrahlung emission originating from the accretion column consistently visible above the WD surface.

Phase-resolved spectroscopy was also performed on the \nicer\ observation taken concurrently with the \nustar\ observation.  However, the plasma temperature estimates were poorly constrained as a result of low exposure time in each of the phases, as well as the fact that the plasma temperature estimated by \nustar\ is above the energy range of the \nicer\ spectrum.  \nicer\ phase-resolved spectroscopy did not suggest an orbital phase dependence in the plasma temperature.

\section{White dwarf mass measurement} \label{sec_mass}

The accretion column is characterized by a radiatively cooling flow of the free-falling partially ionized gas which exhibits varying plasma temperature and density from the stand-off shock to the WD surface. As shown in recent X-ray studies of IPs, \nustar\ hard X-ray spectral data above 10 keV are crucial to constrain the shock temperature, which is directly related to the WD mass \citep{Hailey2016, Suleimanov2019, Shaw2020}. In this section, we describe our X-ray spectral model for polars named {\tt MCVSPEC} and our fitting results to the 3--50 keV \nustar\ spectra. The {\tt MCVSPEC} model has been recently applied to X-ray spectra of IPs \citep{vermette2023constraining, Salcedo2024, Mondal2022}. This paper presents the first application of {\tt MCVSPEC} to the X-ray spectrum of a polar. 

\subsection{X-ray spectral model description} \label{subsec_MCVSPEC}

We employed the {\tt MCVSPEC} model by considering the accretion flow along the magnetic field lines, the plasma temperature and density gradients, and the varying X-ray emissivity along the accretion column. Two different versions of the model have been developed for IPs and polars. 
The polar model version assumes that the free-falling gas begins gaining kinetic energy immediately after leaving the Lagrange point of the Roche lobe of the secondary donor star. The ballistic accretion flow eventually becomes trapped and channeled along the WD magnetic field lines as shown in Doppler tomography observations \citep{schwope1997phase}. Unlike IPs with truncated accretion disks,  it is valid to assume that gas is free-falling from infinity to the stand-off shock height ($h_s$), where the accretion flow becomes supersonic. The free-fall velocity at the shock height is thus $v_{\rm ff} = \sqrt{\frac{2 G M}{R + h_s}}$ where $M$ and $R$ are the WD mass and radius, respectively. For a given WD mass, $R$ can be derived using the well-known WD mass-radius relation \citep{nauenberg1972analytic}. 
At the stand-off shock, the accreting gas is heated to  $kT_s = \frac{3}{8} \frac{G M \mu m_H}{R+h_s}$ where $m_H$ is the hydrogen atomic mass and $\mu$ is the mean molecular weight. Below the shock height, the accretion gas keeps falling toward the WD surface while cooling radiatively by emitting cyclotron optical/UV radiation and thermal X-ray photons. It is especially important to consider the cyclotron cooling process for polars due to their higher WD B-fields compared with IPs.

Some of the X-ray photons emitted from the accretion column may be reflected off the WD surface. The reflected X-rays often appear as a fluorescent neutral Fe line at $6.4$ keV and a Compton hump above 10 keV \citep{hayashi2018x}. While a Gaussian line model is fit to the 6.4 keV neutral Fe line, X-ray continuum reflection is often modeled by \texttt{reflect} in XSPEC \citep{magdziarz1995angle}. We have implemented the reflection model within {\tt MCVSPEC}. The reflection model employs a reflection scaling factor ($r_{refl}$), inclination angle ($i$), and abundance ($Z$) as input parameters.  The abundances in the {\tt APEC} and {\tt reflect} models are linked in the code.  {\tt MCVSPEC} internally calculates the reflection scaling factor defined by $r_{refl} = \Omega/2 \pi = 1-\sqrt{1-\frac{1}{(1+h_s/R)^2}}$ \citep{tsujimoto2018suzaku} for an updated $h_s$ value in each spectral fit iteration. Note that hard X-ray photons ionizing Fe K-shell electrons are emitted from near the shock height. 

The input parameters of the polar {\tt MCVSPEC} model are WD mass ($M$), magnetic field strength ($B$), bolometric luminosity ($L$), fractional accretion area ($f$), abundance of the accretion column and WD surface relative to the solar value ($Z$), inclination angle ($i$) and flux normalization.  It is important to note that the inclination angle used by the {\tt reflect} model, which is implemented in {\tt MCVSPEC}, is not the same as the orbital inclination of the system, as the accretion pole of EF Eri is offset from the spin axis.  As mentioned previously, the orbital inclination has been measured as $i_{\rm{orbital}} = 58^\circ$ \citep{cropper1990polars, campbell2008cyclotron}.  Considering a magnetic colatitude of $17.5^\circ \pm 0.1^\circ$, we take the flux-weighted average of the accretion pole inclination angle across the orbital period to derive a mean value of $47.5^\circ$.  To align with the terminology used in the {\tt reflect} model description, we will refer to this mean accretion pole inclination angle simply as the inclination angle. 
  
For a given WD mass $M$ and bolometric luminosity $L$, the general accretion rate is calculated internally to {\tt MCVSPEC} using the relation $\dot{M} = \frac{L}{G M/R}$.  {\tt MCVSPEC} then calculates a specific accretion rate $\dot{m}$ using the fractional accretion area $f$.  For a given set of $M$, $B$, and $\dot{m}$, coupled differential equations for the flow continuity and momentum/energy conservation are numerically solved along the accretion column to determine the density and temperature profiles in a 1-temperature formulation.  Then, an X-ray model spectrum is calculated by integrating the emissivity (with different $kT$ and $\rho$ values) throughout the accretion column by implementing the most updated atomic database (\url{ http://atomdb.org}) for collisionally ionized plasma (\texttt{APEC}).  As described above, the \texttt{reflect} model is implemented within the {\tt MCVSPEC} model to account for the Compton hump above 10 keV. 
 Since the reflection scaling factor is dependent on the shock height $h_s$ (which is calculated in each iteration), the {\tt MCVSPEC} model will continue to iterate until the shock height produced by the differential equations is consistent with the reflection scaling factor $r_{refl}$ that models the Compton hump in the spectral data.  The {\tt MCVSPEC} model is fully implemented in XSPEC and verified against a handful of mCVs with independent WD mass measurements (Bridges et al. in preparation).  
Hence, our baseline spectral model for polars is {\tt tbabs*(MCVSPEC+gauss)}, which takes into account both the primary accretion column and secondary X-ray reflection emission self-consistently. Similar to the phenomenological models, the {\tt tbabs} and {\tt gauss} models account for the ISM absorption and neutral Fe fluorescent line at 6.4 keV, respectively.

\subsection{X-ray spectral fitting procedures with the \texttt{MCVSPEC} model} \label{subsec_MCVSPECfits}

\begin{figure}[ht] 
    \centering
    \includegraphics[width=1.0\textwidth]{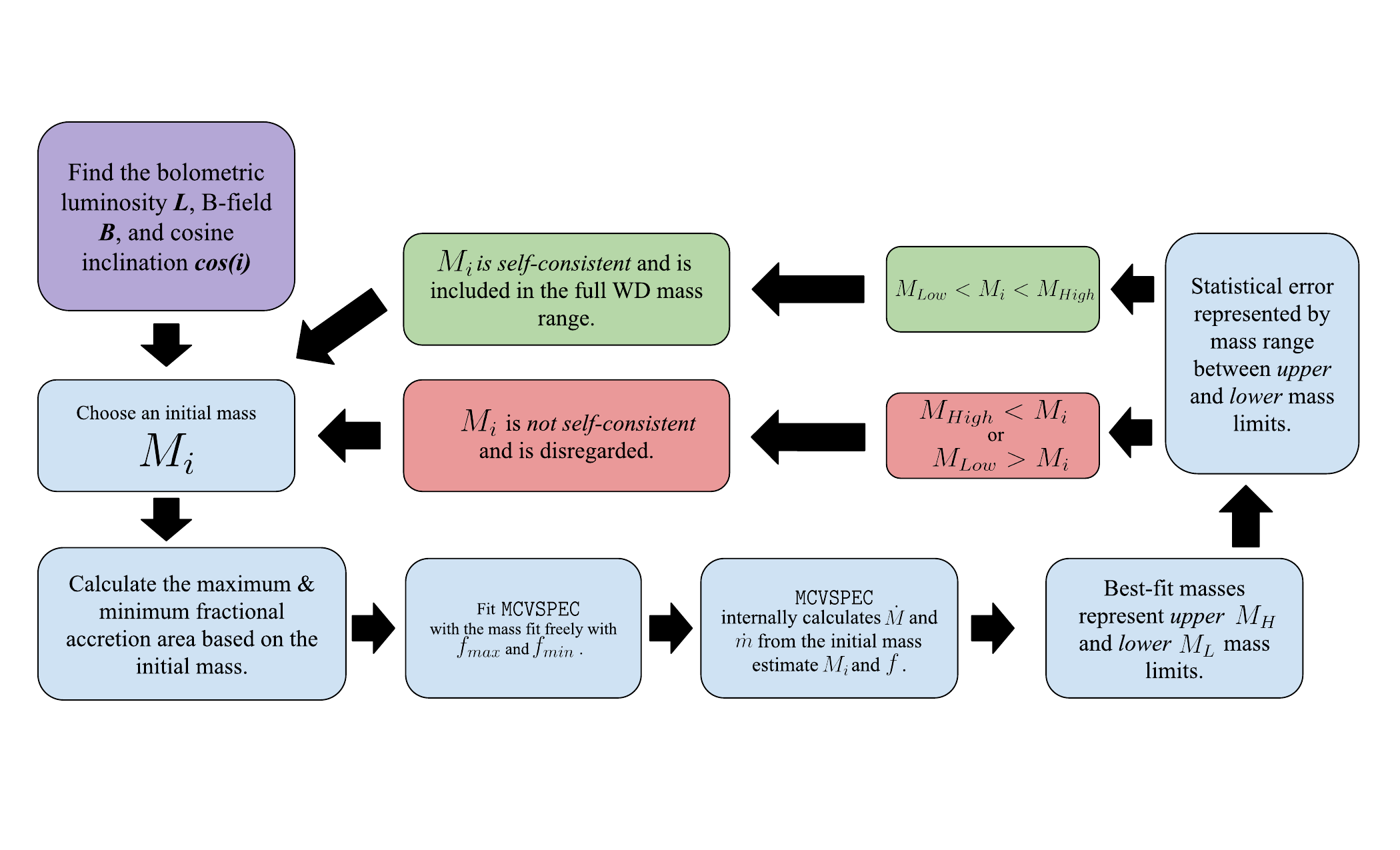} 
    \caption{A flow chart of our iterative procedures of finding a self-consistent solution and deriving the WD mass of EF Eri.  The process begins with inputting the bolometric luminosity, independently measured B-field strength and cosine inclination (\textcolor{violet}{purple}).  The main sequence of running MCVSPEC (\textcolor{cyan}{blue}) consists of fitting the X-ray spectrum with multiple initial mass assumptions.  After running MCVSPEC, if the best-fit mass is in agreement with the initial mass assumption (\textcolor{green}{green}), that initial mass $M_i$ is considered to be self-consistent with the constraints on $f$ and \mdot\ that it demands (See \S5.2.2, \S5.2.3) and is included in the final WD mass range.  If the best-fit mass is not in agreement with the initial mass $M_i$ (\textcolor{red}{red}), the initial mass estimate is excluded from the final WD mass range.
    }
    \label{flow_chart}
\end{figure}

Among the input parameters for {\tt MCVSPEC}, we set the inclination angle to be $i = 47.5^\circ$, and the magnetic field strength $B$ to be $13$ MG \citep{Campbell2008}.  In the case of EF Eri, WD mass $M$, fractional accretion area $f$, abundance $Z$, and flux normalization are fit parameters.    
The specific mass accretion rate $\dot{m}$ is linked to $f$ through $\dot{m} = \frac{\dot{M}}{4 \pi {R}^2f}$. 
When $\dot{m}$ is large, the accretion column height becomes shorter ($h_s \ll R$). In this case, both the shock temperature and X-ray reflection become nearly independent of $h_s$ because $kT_s \propto \frac{M}{R}$ and $\Omega/2\pi \approx 1$. 
On the other hand, when  $\dot{m} \simlt 1$ g\,cm$^{-2}$\,s$^{-1}$, $h_s$ increases and alters the X-ray spectra significantly from the short accretion column case \citep[$h_s \ll R$; ][]{Hayashi2014, Suleimanov2016}.  
Hence, when $\dot{m}$ is smaller, the WD mass measurement is subject to larger systematic errors which depend on the (unconstrained) $\dot{m}$ value. For this reason, our first step is to constrain a range of $\dot{m}$ based on the bolometric luminosity (\S5.2.1) and fractional accretion column area (\S5.2.2).  Since $\dot{M}$ and $\dot{m}$ also depend on $M$, we assume an initial WD mass value ($M_i$) and fit {\tt MCVSPEC} to the \nustar\ spectra for the derived $\dot{m}$ range (\S5.2.3). Once this second step is completed, we check whether the best-fit WD mass ($M_f$) is consistent with $M_i$ within statistical and systematic errors. We step through different $M_i$ values, perform a spectral fit, and validate the self-consistency of each $M_i$ (\S5.2.4). Our final results are obtained through these iterative processes as summarized in a flow chart (Figure \ref{flow_chart}).

\subsubsection{Estimating bolometric luminosity}

The bolometric luminosity in polars consists of three components: (1) bremsstrahlung emission $L_{bremss}$ in the X-ray band ($E\simgt 1$ keV), (2) cyclotron emission $L_{cyc}$ in the IR, optical, and UV bands, and (3) blackbody radiation $L_{\rm BB}$ in the soft X-ray band ($E\simlt1$ keV). 

\begin{itemize} 
\item[(1)] The unabsorbed X-ray flux measured from the \nustar\ observation is ($1.15 \pm 0.01) \times {10}^{-10}$ \fluxcgs\ ($0.1 \rm{-} 100$ keV). Although the \nustar\ energy range is 3--50 keV for spectral fitting, we extrapolated the X-ray flux to 0.1--100 keV in XSPEC and estimated the thermal X-ray luminosity.  The unabsorbed X-ray flux measured from the \nicer\ observation was in agreement at ($1.32 \pm 0.18) \times {10}^{-10}$ \fluxcgs\ when extrapolated over the same 0.1--100 keV energy range.  Taking uncertainty weighted average of the two estimates we arrive at an estimated flux of ($1.15 \pm 0.01) \times {10}^{-10}$ \fluxcgs\ due to the superior accuracy of the \nustar\ estimate.  Considering the source distance of $160 \pm 4$ pc derived from \textit{Gaia} parallax measurements, we find $L_{bremss} = (1.76\pm0.1) \times {10}^{32}$ \lumcgs.  

\item[(2)] To estimate the cyclotron luminosity, we compared the low-state fluxes derived from GALEX-UV and optical BVRI photometry to the high-state flux derived from ASAS-SN g-band optical data taken hours before the \nustar\ observation \citep{szkody2006galex}.  The optical V-band flux of $\sim 1.0 \times {10}^{-12}$ \fluxcgs\ observed during the low state comprised about $\sim 20\%$ of the total IR-optical-UV flux of $5.7 \times {10}^{-12}$ \fluxcgs\ \citep{harrison2003modeling}.  Using the g-band flux of $9.5 \times {10}^{-12}$ \fluxcgs\ derived from the ASAS-SN data during the current high state and extrapolating based on the UV-BVRI photometry data, we estimate the high state cyclotron flux to be $F_{cyc} = 4.7 \times {10}^{-11}$ \fluxcgs.  This estimate is in close agreement with the expected high state cyclotron flux $4.8 \times {10}^{-11}$ \fluxcgs\ derived during EF Eri's previous high state \citep{schwope2007xmm, beuermann1987einstein}.  The luminosity of the cyclotron component is thus $L_{cyc} = 7.2\times10^{31}$ \lumcgs.

\item[(3)] Among the previous reports on detecting the blackbody radiation from the emission region of EF Eri, we selected the \textit{EUVE} observations performed in 1993 \citep{1999ASPC..157..157M}.   The blackbody flux derived from the \textit{EUVE} observations during the previous high state was ${5.1}^{+5.9}_{-2.1}\times10^{-11}$ \fluxcgs.  Given the similarity in the bremsstrahlung and cyclotron luminosities between the current and pre-1997 high states, we assume that the current blackbody luminosity is similar to that recorded by the \textit{EUVE} observation during the previous high state.  Using the same distance of $160 \pm 4$ pc we find $L_{\rm{BB}} = {7.8}^{+9.0}_{-3.2} \times {10}^{31}$ \lumcgs\ . 

\end{itemize}

Combining all components $L_{bremss}$, $L_{cyc}$, and $L_{\rm{BB}}$ we estimate the bolometric luminosity of EF Eri in the high state to be $L = ({3.26}^{+0.9}_{-0.3})\times{10}^{32}$ \lumcgs. Since the estimated $L$ value is significantly lower than the Eddington luminosity limit ($\sim  7\times {10}^{37}$ \lumcgs), the effect of radiation pressure on accretion column structure is negligible \citep{wolff1989stability}.  

\subsubsection{Constraining fractional accretion column area} \label{Mass_f_relation}
  
While $f$ is largely unknown for most IPs and polars, there are some mCVs (including EF Eri) where $f$ has been constrained observationally. 
Since the accretion column geometry of EF Eri is not known, we adopt the next best approach of constraining it using observational data.  In the case of EF Eri, the fractional accretion area $f$ is most robustly constrained by detecting blackbody radiation from the WD polar caps.  The blackbody radiation is considered as arising from thermalized particles in the WD atmosphere bombarded by in-falling gas along the accretion column \citep{kuijpers1982comments, woelk1996stationary, litchfield1990blob}. Critically, this method only provides a maximum limit for $f$ as the area being heated by accreting material may be larger than the area upon which the gas is accreted directly.   The \nicer\ observation taken concurrently with the \nustar\ observation is essential for this analysis as the $3.0 \rm{-}79$ keV energy range of \nustar\ precludes the analysis of soft X-ray features.

The latest and most accurate measurement of the blackbody temperature of EF Eri's emission region is provided by the \textit{EUVE} observation taken in 1993.  \citet{1999ASPC..157..157M} found evidence of a significant blackbody emission feature with $kT_{BB} = 19.4 - 30.4$ eV present in the $88-176$ eV energy range available to the \textit{EUVE} instrument.   Given that our \nicer\ spectrum does not cover the energy range of the previously observed blackbody component, we are not able to directly constrain the blackbody temperature using the \nicer\ data.  Therefore, we follow the method used by \citet{worpel2017x} to constrain the blackbody temperature and normalization of a similar polar system V808 Aur.

We attempt to determine what blackbody component could be present but unobserved in the energy region below the \nicer\ spectrum.  Our reasoning is as follows: if there was such a blackbody component, its high energy tail should be apparent in the \nicer\ soft X-ray spectrum.  The influence of a blackbody component on the soft X-ray fit increases with $kT_{BB}$ and the normalization parameter $K$.  The normalization of the \texttt{bbodyrad} model is proportional to the radius of the accretion region.  Since our purpose is to constrain the maximum possible accretion area, we need to find the maximum possible normalization $K$.  The highest possible normalization factor that does not degrade the soft X-ray \nicer\ fit will thus be found at the lowest possible $kT_{BB}$, which we take to be $19.4$ eV from the 1993 \textit{EUVE} observation.

\begin{figure}[ht] 
    \centering
    \begin{minipage}[b]{0.49\textwidth}
        \includegraphics[width=\textwidth]{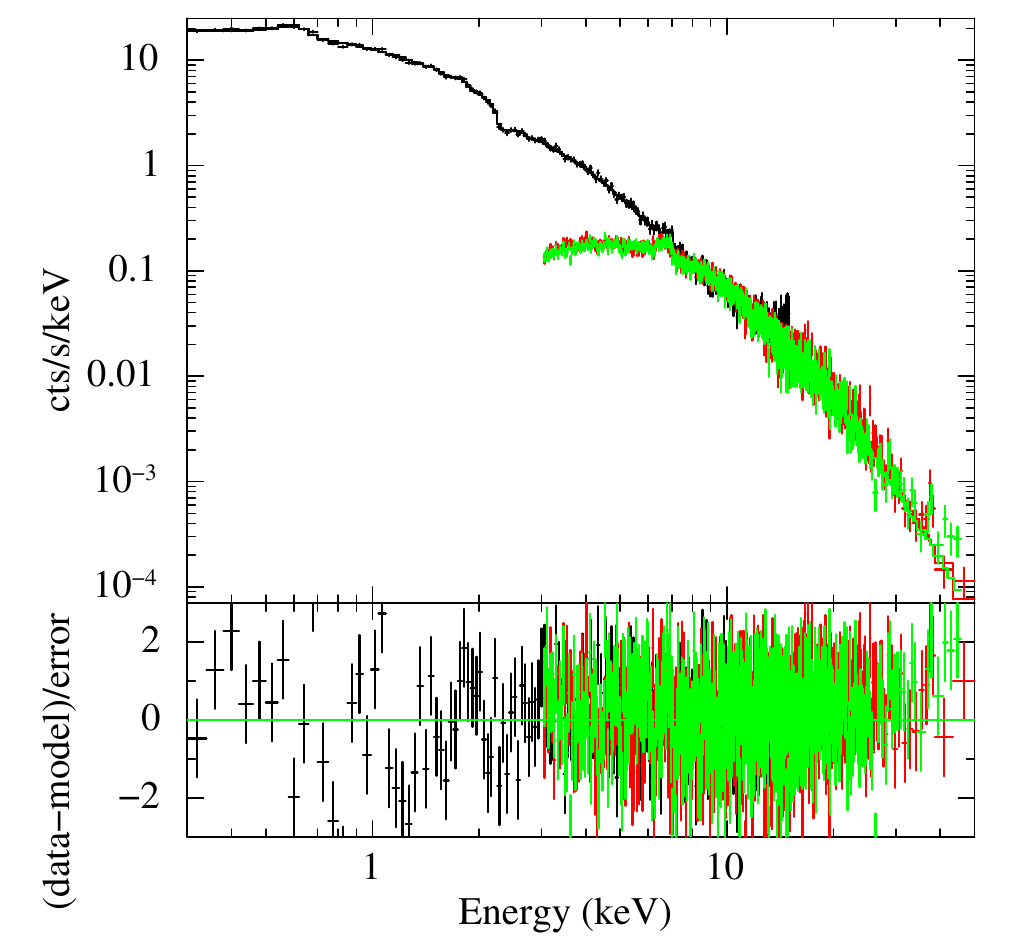}
    \end{minipage}
    \hfill
    \begin{minipage}[b]{0.49\textwidth}
        \includegraphics[width=\textwidth]{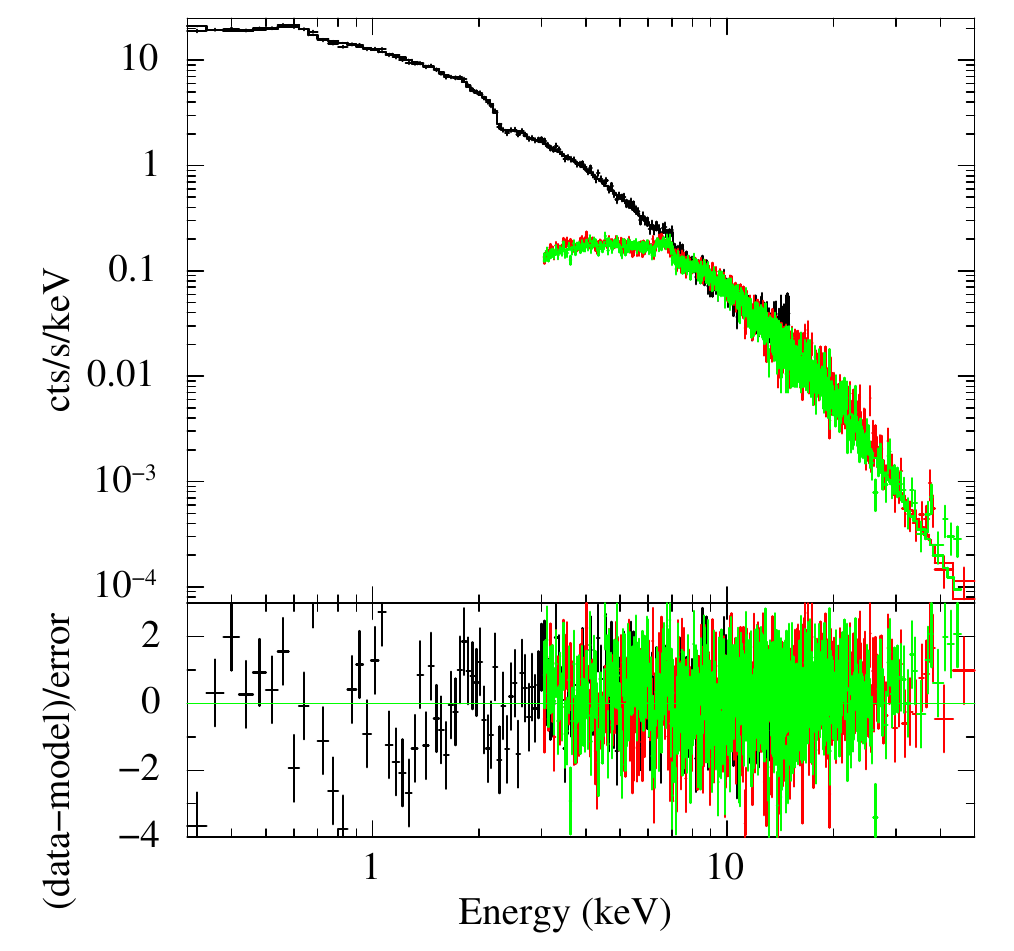}
    \end{minipage}
    \caption{Left: Joint $\nustar$ and \nicer\ spectra and residuals of EF Eri fit with a \texttt{tbabs*(APEC+bbodyrad+gauss+gauss)} model, yielding a $\chi^2$ statistic of $0.99$ ($1054$ d.o.f.).  Right: Joint $\nustar$ and \nicer\ spectra and residuals of EF Eri fit with a \texttt{tbabs*(APEC+bbodyrad+gauss+gauss)} model, yielding a $\chi^2$ statistic of $1.00$ ($1054$ d.o.f.).  The $\nicer$ observation is in black, presented in the 0.3--12 keV band.  The $\nustar$ observation is presented in the 3--50 keV band, with FPMA in red and FPMB in green.
    }
    \label{blackbody_joint}
\end{figure}

Figure \ref{blackbody_joint} presents two joint fits with \nustar\ and \nicer\ observations.  The complete model used in this fit is \texttt{const*tbabs*(APEC+bbodyrad+gauss+gauss)}.  The first gaussian component accounts for the neutral iron K$\rm{-}\alpha$ emission feature that is prevalent in both observations.  The second gaussian component exists to account for an OVII photoionization line that appears at 0.59 keV. 
The presence of an OVII photoionization line has been previously recorded in polars (e.g., \citealt{ramsay2004xmm}), however there is a possibility that the photoionization line is a NICER background artifact resulting from solar X-rays ionizing neutral oxygen in the earth's polar cusp regions.\footnote[1]{SCORPEON Background Model Overview can be found at: \href{https://heasarc.gsfc.nasa.gov/docs/nicer/analysis\_threads/scorpeon-overview/}{https://heasarc.gsfc.nasa.gov/docs/nicer/analysis\_threads/scorpeon-overview/}}  Regardless, the feature at 0.59 keV is fit well by a simple gaussian component and does not influence the results of our blackbody fitting.

In Figure \ref{blackbody_joint} (left panel), the blackbody temperature is fixed to the lowest $kT_{BB}$ estimate from the 1993 \textit{EUVE} observation.  We utilize the $\Delta \chi^2$ between this best fit case and other cases with normalization $K$ fixed above its best-fit value.  Figure \ref{blackbody_joint} (right panel) represents a $\Delta \chi^2$ of 9, analogous to a $3\sigma$ probability of occurrence when varying one degree of freedom (normalization $K$).  The normalization $K$ associated with Figure \ref{blackbody_joint} (right panel) is $5 \times {10}^{8}$, corresponding to a maximum fractional accretion area of $4.4 \times {10}^{-4}$ with an initial mass estimate $M_i = 0.6 M_\odot$

However, the maximum $f$ value will increase if the initially assumed WD mass $M_i$ increases, given that the surface area of the WD decreases.  For instance, the maximum $f$ value for a 1.3$M_\odot$ WD using this emission area is $f_{\rm max} = 5.3 \times {10}^{-3}$, while for a 0.4$M_\odot$ WD it is $f_{\rm max} = 2.8 \times {10}^{-4}$.  

\subsubsection{WD mass dependence on the specific accretion rate} \label{Mass_mdot_relation}

\begin{figure}[ht] 
    \centering
    \begin{minipage}[b]{0.49\textwidth}
        \includegraphics[width=\textwidth]{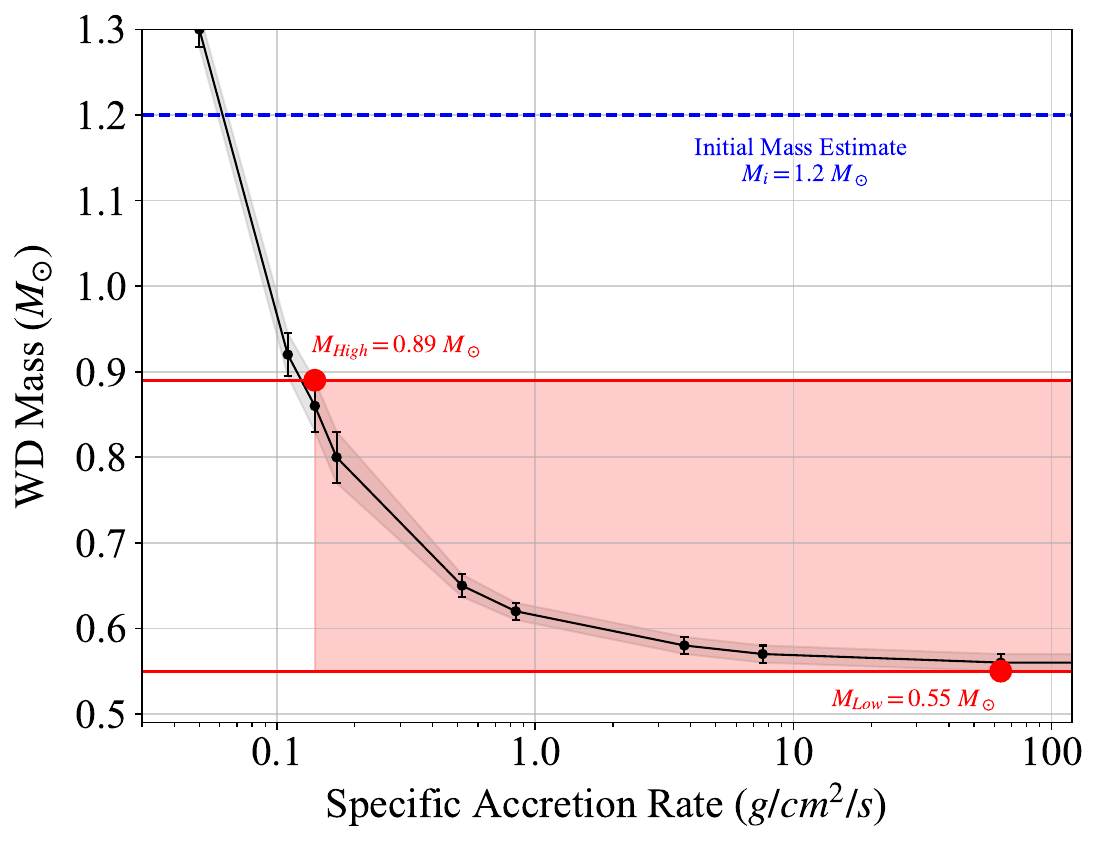}
    \end{minipage}
    \hfill
    \begin{minipage}[b]{0.49\textwidth}
        \includegraphics[width=\textwidth]{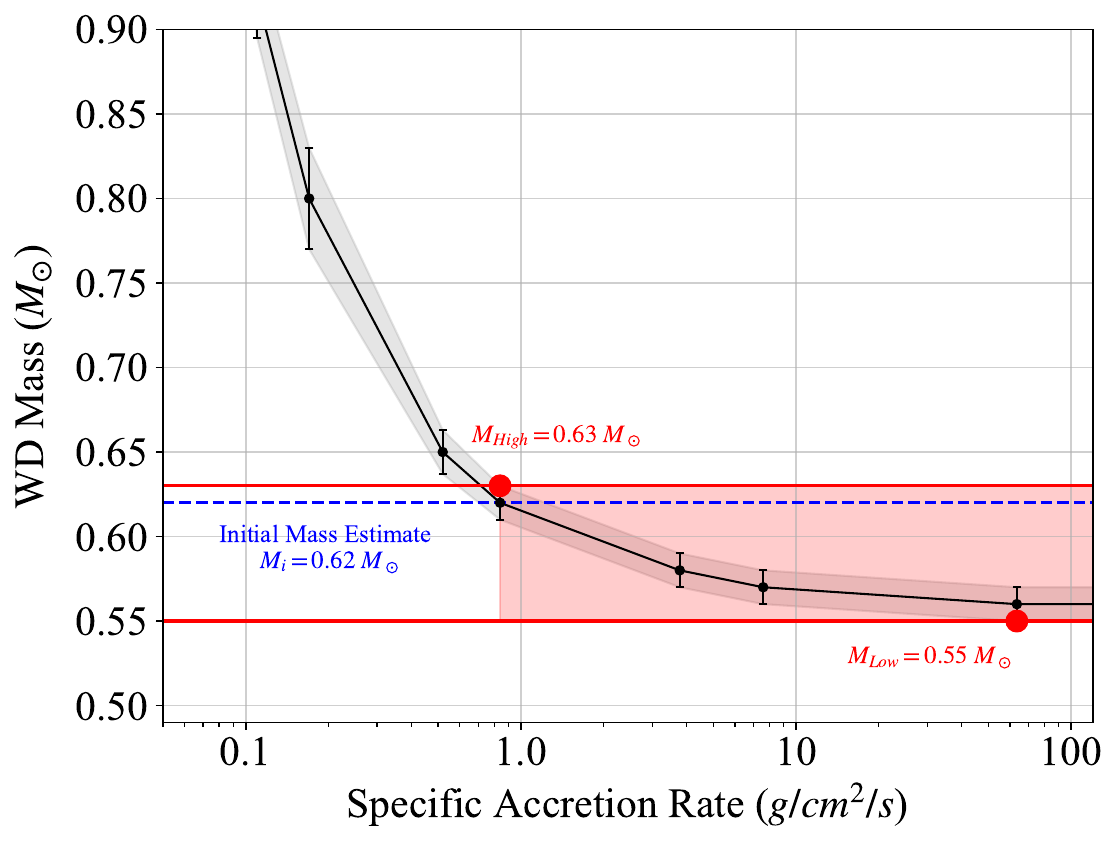}
    \end{minipage}
    \caption{$\dot{m}$ vs. the best-fit WD mass ($M_f$) plots for $M_i = 1.2 M_\odot$ (left) and $0.62 M_\odot$ (right) derived from fitting $\nustar$ spectra of EF Eri.  The red shaded area corresponds to the best-fit mass range, through fitting the {\tt MCVSPEC} model to the \nustar\ spectra, within the valid $\dot{m}$ ranges. The blue dashed lines indicate the initial WD mass values assumed for the spectral fitting. The left and right plots represent the inconsistent and self-consistent cases, respectively, as described in the main text.
    }
    \label{fig_mdotvmass1.0}
\end{figure}

Before we derive a self-consistent solution for the WD mass measurement, here we present how the best-fit WD mass varies with $\dot{m}$, assuming $M_i = 1.2 M_\odot$. In this case, we derive the maximum fractional accretion area of $f_{\rm max} = 2.5 \times {10}^{-3}$ and $\dot{M} = 7.3 \times {10}^{14}$ g\,s$^{-1}$.  Since $f_{\rm max}$ corresponds to the minimum possible specific accretion rate, we calculate the minimum specific accretion rate to be $\dot{m} = 0.18$ g\,cm$^{-2}$\,s$^{-1}$.  
We then fit the {\tt MCVSPEC} model to the 3--50 keV \nustar\ spectra for multiple $\dot{m}$ values between $0.1 \rm{-} 120$ g\,cm$^{-2}$\,s$^{-1}$.  In each spectral fit, we recorded the best-fit WD mass ($M_f$) and its statistical error.  Figure \ref{fig_mdotvmass1.0} (left panel) presents a \mdot\ vs $M_f$ relation for $M_i = 1.2 M_\odot$.   
The $M_f(\dot{m})$ curve is characterized by a rapid variation in $M_f$ at lower $\dot{m}$ values and flattening above $\dot{m}\sim 1$ g\,cm$^{-2}$\,s$^{-1}$.   The flattening is due to the short accretion column height ($h_s \ll R$), in which case the shock temperature and X-ray reflection become nearly independent of $h_s$ or $\dot{m}$. In this mass saturation regime, when $\dot{m}$ is high, the systematic error in WD mass measurement is greatly reduced. 
In the plot, the red shaded region represents the allowed specific accretion rate range of $\dot{m} \ge 0.2$ g\,cm$^{-2}$\,s$^{-1}$.
In this case, the valid $\dot{m}$ range causes the fitted WD mass to be constrained to $M_f \le 0.89 M_\odot$. To be more precise, fitting the \nustar\ spectra by fixing $\dot{m} = 0.2$ g\,cm$^{-2}$\,s$^{-1}$ results in $M_f = 0.86 \pm 0.03 M_\odot$.  The upper bound corresponds to $M_f \le 0.89 M_\odot$. Picking an arbitrarily large specific accretion rate at  $\dot{m} = 120$ g\,cm$^{-2}$\,s$^{-1}$ yielded $M_f = 0.56 \pm 0.01 M_\odot$.

As is evident in Figure \ref{fig_mdotvmass1.0} (left panel), the initial mass estimate of $M_i = 1.2 M_\odot$ is not within the best-fit mass range of $ 0.55 M_\odot \le M_f \le 0.89 M_\odot$.  Therefore, the initial mass estimate of $M_i = 1.2 M_\odot$ should be ruled out.  

\subsubsection{The self-consistent WD mass measurement}

\begin{figure}[ht] 
    \centering
    \includegraphics[width=0.6\textwidth]{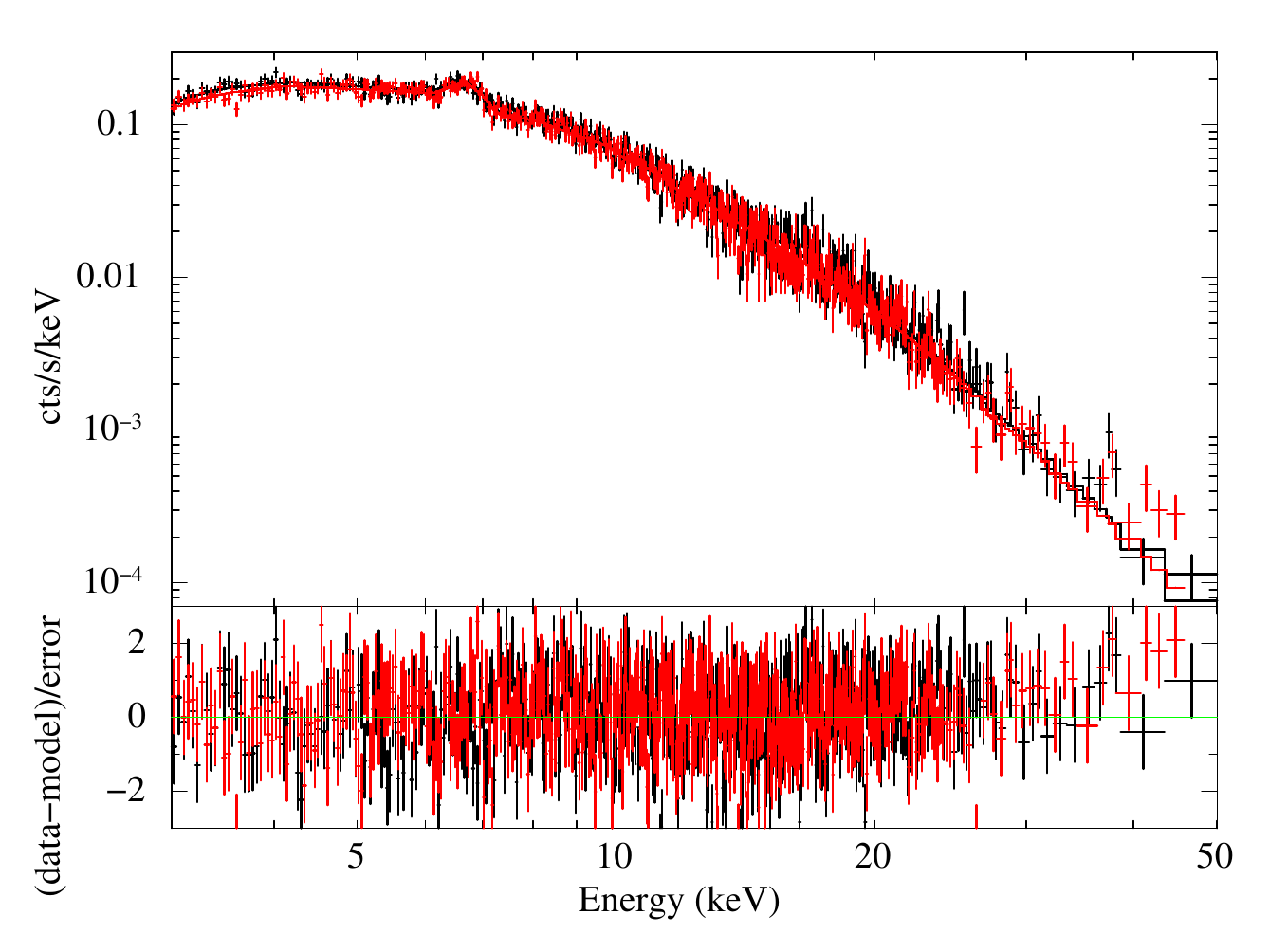} 
    \caption{The $\nustar$ spectra and residuals fit with the {\tt MCVSPEC} model with $M_i = 0.62 M_\odot$ and \mdot = 0.84 g\,cm$^{-2}$\,s$^{-1}$  (corresponding to $f = 4.6 \times {10}^{-4}$). The best-fit WD mass is $M_f = 0.62 \pm 0.01 M_\odot$, making the fit self-consistent as $M_f \approx M_i$. The other fit parameters are listed in Table \ref{tab:mcvspec_fit} ($\dot{m}_{\rm{min}}$ Case).}
    \label{fig_MCVSPEC}
\end{figure}

We repeated the same spectral fitting procedures for other $M_i$ values between $0.5 M_\odot$ and $1.2 M_\odot$. First, we incremented $M_i$ by $0.1M_\odot$ and later we fine-tuned our fitting results using a smaller grid size ($\Delta M_i = 0.01 M_\odot$) around a self-consistent solution candidate. For each $M_i$ value, we validated the self-consistency by imposing the sufficient condition of $M_i \approx M_f$ within statistical and systematic errors.  As shown in the previous section, the systematic error depends on $\dot{m}$, and it is greatly reduced when the $\dot{m}$ range is in the mass saturation regime. Figure \ref{flow_chart} illustrates a flowchart describing our methodology to determine whether an initial mass estimate is self-consistent and subsequently hould be included in the final WD mass range. As discussed in the previous section, $M_i = 1.2 M_\odot$ represents one of the self-inconsistent cases. We found the only self-consistent initial mass estimates to be $M_i = (0.55-0.63) M_\odot$. For instance, considering $M_i = 0.62 M_\odot$, we calculated  $f_{\rm max} = 4.6 \times {10}^{-4}$ and $\dot{M} = 3.3 \times {10}^{15}$ g\,s$^{-1}$.  The $M_f(\dot{m})$ curve for $M_i = 0.62 M_\odot$ is presented in Figure \ref{fig_mdotvmass1.0} (right panel), with  $\dot{m} \ge$ 0.84 g\,cm$^{-2}$\,s$^{-1}$, yielding excellent fits to the \nustar\ spectra. In Table \ref{tab_MCVSPEC}, 
we present spectral fitting results for two extreme $\dot{m}$ values of 0.84 and 120 g\,cm$^{-2}$\,s$^{-1}$ corresponding to the lower bound and an arbitrarily high accretion rate well within the mass saturation regime.  As shown in Table \ref{tab_MCVSPEC}, the best-fit mass with statistical error for the \mdot\ = 0.84 g\,cm$^{-2}$\,s$^{-1}$ case is $0.62 \pm 0.01 M_\odot$, while for the \mdot\ = 120 g\,cm$^{-2}$\,s$^{-1}$ case the best-fit mass is $0.56 \pm 0.01 M_\odot$.   The fitted \nustar\ spectra and residuals for the $\dot{m} = 0.84$ g\,cm$^{-2}$\,s$^{-1}$ case are presented in Figure \ref{fig_MCVSPEC}. 
The systematic error due to varying $\dot{m}$ is greatly reduced because the possible specific accretion rates are largely within the saturation regime.  Finally, we determined the WD mass to be $M = (0.55 \rm{-} 0.63)$\ms\ after taking into account both the statistical and systematic errors.

\begin{deluxetable*}{lcc}[ht] \label{tab_MCVSPEC}
\tablecaption{{\tt MCVSPEC} fit parameters to the \nustar\ spectra of EF Eri assuming $M_i = 0.57 M_\odot$.}
\label{tab:mcvspec_fit}
\tablecolumns{3}
\tablehead{
\colhead{Parameter}   
&
\colhead{$\dot{m}_{\rm{min}}$ Case}  
& 
\colhead{$\dot{m}_{\rm{high}}$ Case}
}
\startdata  
$N^{(i)}_H [10^{19} \rm{cm}^{-2}]^a\;*$ & $1.0$ & $1.0$ \\
$f$ [g\,cm$^{-2}$\,$s^{-1}$]* & $4.6 \times {10}^{-4}$ & $3.1 \times {10}^{-6}$ \\
$\dot{m}$ [g\,cm$^{-2}$\,$s^{-1}$]* & $0.84$ & $120$ \\
$M [M_{\odot}]$ & $0.62 \pm 0.01$   & $0.56 \pm 0.01$  \\
$Z^b (Z_\odot)$ & ${0.21}^{+0.03}_{-0.04}$  & $0.20 \pm 0.03$ \\
$EW_{\rm line}$ (eV) & $45.6^{+14.7}_{-16.2}$  & $34^{+20}_{-15}$ \\
$refl$* & $0.69$  & $0.98$ \\
$cos(i)$* & $0.68$ & $0.68$  \\
$\chi^2_{\nu}$ (dof) & 0.95 (897)  & 0.97 (897) \\
$h_s/R_{WD}$ & 6.9\%   & 0.04\%  \\
\enddata
$^a$ The ISM hydrogen column density is associated with {\tt tbabs}, which is multiplied to all the models. Due to the $3 - 79$ keV energy range of \nustar, the ISM hydrogen column density has minimal effect on the spectral fit.  $N_H$ is thus frozen to the low column density inferred from \rosat\ and {\textit EXOSAT}  \citep{beuermann1991short, watson_exosat}.\\
$^b$ Abundance relative to solar. \\
$^*$ The parameter is frozen. \\
\end{deluxetable*}

\section{Discussion} \label{sec_disc}

In this section, we will present the conclusions from our analysis of \nustar\ timing and spectral data of EF Eri. In \S\ref{subsec_timingdisc}, we summarize our findings from the \nustar\ timing analysis and discuss the implications of the non-detection of QPOs. In \S\ref{subsec_spectradisc}, we discuss the broadband X-ray properties from phase-averaged and phase-resolved spectral analysis, compare our results with previous WD mass measurements, and highlight the importance of hard X-ray observation in the study of polars.

\subsection{X-ray timing properties} \label{subsec_timingdisc}

Using the Lomb-Scargle algorithm, we detected the known orbital period of EF Eri ($P_{orb} = 81$ min) up to 50 keV.  The X-ray pulse profiles throughout 3--50 keV are characterized by a sharp drop at phase $\sim$ 0.3 and a slow rise between phase 0.7 and 1. 
EF Eri's characteristic dip in soft X-ray and IR luminosity, suggested to be the accretion stream eclipsing the accretion region on the WD surface, is absent in the hard X-ray \nustar\ lightcurves \citep{beuermann1987einstein}, as expected. Since the X-ray lightcurve shapes are nearly energy-independent, we assumed that the modulation was caused largely by the phase-dependent visibility of the accretion column. By fitting a fiducial accretion column visibility model to the \nustar\ lightcurve data, we derived the magnetic colatitude of 17.5\textdegree $\pm$ 0.1\textdegree, which helped to constrain X-ray reflection off the WD surface in our spectral analysis.  In future work, soft X-ray ($E < 3$ keV) lightcurve data utilizing \nicer\ data may provide additional details on EF Eri's accretion column structure.

In the \nustar\ timing data, we found no X-ray QPO signal at $\nu =  0.1\rm{-}100$ Hz (\S\ref{sec_pds}). The previous \xmm\ observations of EF Eri constrained an X-ray QPO amplitude to $A < 58$\% at $\nu = 0.3\rm{-}0.6$ Hz where the optical QPOs have been detected \citep{VanBoxSom2017}. The \nustar\ timing data allowed us to put a significantly more stringent constraint of $A < 5$\% at $\nu = 0.5$ Hz. However, the optical QPOs from EF Eri exhibited lower amplitudes at $A = 1\rm{-}1.3$\% \citep{larsson1987discovery}. It is possible that an X-ray QPO signal at the same frequency with similarly low amplitude may be present but slipped the detection. 

The optical and X-ray QPOs are relevant to the stability conditions of the accretion column. Since the 1980s, it has been theoretically recognized that magnetically-confined accretion flow is inherently unstable if the radiative cooling is dominated by thermal bremsstrahlung radiation \citep{langer1981thermal, langer1982time, Imamura1984}. The thermal instability causes the shock to oscillate on the bremsstrahlung cooling time scale $t_{\rm br} \sim 3 (kT/10\,\rm{[keV]})^{1/2}  (n/10^{15}\,\rm{cm}^{-3})^{-1}$ sec. The shock oscillation is predicted to be a universal phenomenon in polars \citep{Wu2000, wolff1991noise}, NS-HMXBs \citep{Sheng2023} and T-Tauri stars \citep{Drake2009}, and should be manifested as QPOs. For polars, optical and X-ray QPOs at $\nu\sim1$ Hz have been predicted by numerous analytical and simulation studies. So far, despite the optical band detection of 1--3-sec QPOs from 5 polars, no detections of X-ray QPOs have been reported from the \xmm\ observations of 24 polars \citep{Bonnet2015}. The rarity of optical QPOs and the lack of X-ray QPOs puzzlingly suggest that accretion columns are mostly stable. This is inconsistent with various plasma instability models proposed in the past (e.g., \citealt{Wu2000}). 

Recent 2D radiative MHD simulations predicted X-ray QPOs may appear at higher frequencies ($\nu \sim1\rm{-}100$ Hz) with large amplitudes  \citep[$A\sim60$\%; ][]{VBS2018}. 
The MHD simulations predict higher QPO frequency and larger amplitude when the ratio of thermal bremsstrahlung and cyclotron time (characterized by $\epsilon_s = t_{\rm brem}/t_{\rm cycl}$ at the shock height) $\epsilon_s \simlt 1 $ is the necessary condition to induce the thermal plasma instability within the accretion column. The condition is satisfied for lower WD B-fields and higher specific accretion rates. Given the relatively lower WD B-field of EF Eri $B = 13$ MG within the known range of $B = 7\rm{-}230$ MG, an X-ray QPO associated with the shock oscillation should appear at $\nu\simgt10\rm{-}100$ Hz \citep{VBS2018}. 
For a given WD B-field, both the X-ray QPO frequency and amplitude increase with $\dot{m}$ as a result of the lower $\epsilon_s$ ratio. In particular, when $\dot{m} \simgt 30$~g\,cm$^{-2}$\,s$^{-1}$, the QPO amplitude reaches $\sim50$\%. Such a high specific accretion rate is possible during the \nustar\ observation as we constrained $\dot{m} \simgt 0.84$~g\,cm$^{-2}$\,s$^{-1}$ in the spectral analysis section (\S\ref{subsec_MCVSPECfits}). However, the QPO amplitude upper limit at $\nu = 10$ Hz ($A<140$\%) is higher than the predicted range of $A\simlt50$\%. We need a much higher count rate ($R\gg1$ cts\,s$^{-1}$) to detect the predicted X-ray QPOs at $\nu\simgt10$ Hz. 
Moreover, another MHD simulation study suggested that shock oscillations could be enhanced by magnetic field perturbation at the base of the accretion column \citep{Bera2018}. Accumulated gas near the WD surface distorts magnetic field lines and induces Alfv\'enic wave oscillations. In this case, QPOs could be found at lower energies than the \nustar\ band ($E<3$ keV) as soft X-ray photons are emitted from the bottom of the accretion column.  Therefore, the \nicer\ timing data of EF Eri are best suited to search for the X-ray QPOs predicted at a higher frequency range and/or lower energy band.

\subsection{X-ray spectral properties} \label{subsec_spectradisc}

    The analysis conducted on this \nustar\ hard X-ray observation has provided the most accurate mass estimate for EF Eri to date. Previous mass estimates based on X-ray observations have either overlooked the significant influence of shock height on constraining the WD mass or suffered from poor photon statistics in the hard X-ray regime, due to EF Eri's prolonged low state \citep{beuermann1987einstein, wu1995x, cropper1999effects}.  As discussed in \S\ref{subsec_MCVSPEC}, the shock temperature $kT_s$ and shock height $h_s$ are related via the formula $kT_s = \frac{3}{8} \frac{G M \mu m_H}{R+h_s}$.  Accurate estimation of the shock height is essential to ensure consistency between the observed degree of reflection off the WD surface and the calculated specific accretion rate, $\dot{m}$, which directly impacts the bremsstrahlung emissivity through its effect on plasma density, $\rho$.  Given that polars can exceed shock temperatures of $kT_s \sim 20$ keV during periods of high accretion, the ability of \nustar\ to achieve robust photon statistics above 10 keV is crucial for obtaining accurate mass estimates.

Our WD mass measurement of $(0.55 - 0.63)$\ms, derived by modeling the shock height and accretion column consistently with the observed degree of reflection and specific accretion rate, represents the most reliable mass determination for EF Eri.  Considering the updated distance estimate of $160 \pm 4$ pc from \textit{Gaia} parallax measurements, our results closely align with the WD mass range derived from optical/IR observations \citep{schwope2010x}.  The independent estimate was based on radial velocity measurements corrected for irradiation effects, providing a mass range between $\sim (0.55-0.65)$\ms.

    High accretion states represent a fundamental shift in the emissivity of polars.  Observations of EF Eri in its low accretion state illustrated that cyclotron emission is reduced by a factor of $\sim 10$, while thermal bremsstrahlung emission is reduced by a factor of $\sim 2500$ \citep{schwope2007xmm}.  This results in vastly fewer high-energy X-ray photons being produced, resulting in poor photon statistics and limited constraints, thus increasing statistical error.

    Our identification of the degeneracy between $M$ and \mdot\ at low accretion rates has further increased the need for observations during high accretion states to reduce systematic error.  As shown during our analysis in \S\ref{subsec_MCVSPEC}, predicted $M$ will change rapidly when \mdot\ is low as a consequence of shock height $h_s$ being very sensitive to changes in a low specific accretion rate.  As \mdot\ is increased, the relative height of the shock height compared to the WD radius is greatly reduced, and the mass estimate enters a saturation region with \textit{M} becoming nearly independent of \mdot.  Given the rarity of CVs in which the fractional accretion area can be conclusively determined through deep X-ray eclipses, and thus the uncertainty in \mdot\ can be greatly reduced, observing systems in a high state of accretion allows for more accurate WD mass measurements. Therefore, our findings highlight the importance of observing polars in high accretion states with \nustar\ for two reasons - (1) improving photon statistics (especially at higher energies for constraining the shock temperature) and (2) entering the WD mass saturation region. (1) and (2) reduce the statistical and systematic errors associated with WD mass measurements for polars, respectively. 

    We note that the mass estimate of $(0.55 - 0.63) M_\odot$ we derive for EF Eri is lower than the mean mass of magnetic CVs studied by \cite{Ramsay2000} of $(0.84 \pm 0.20)M_\odot$ and marginally lower than the mean mass of isolated WDs with magnetic field strengths $B  > 1$ MG ($0.66\pm0.14 M_\odot$) \citep{Ferrario2015}. This not only puts EF Eri on the low end of the mass range for magnetic CVs (\citealt{Suleimanov2019, de2020hard}) but also non-magnetic WD CVs \citep{Pala2020, pala2022constraining}. In contrast, it is similar to the peak in the mass distribution of isolated DA WDs ($0.59 M_\odot$) \citep{Kilic2020}. One of our wider goals of the NuSTAR polar observation program is to determine the mass of magnetic white dwarfs in accreting binaries in a statistically significant sample and compare them to the mass of isolated white dwarfs and how that can inform us about their formation and evolution.
 
\section{Summary} \label{sec_summ}

    The first hard X-ray observation of EF Eri in a high state has provided the most accurate mass estimate ever provided for this source, with $M = (0.55 - 0.63)$\ms.  The estimate was produced using the multi-temperature model {\tt MCVSPEC}, which considers the effect of shock height, the ratio between bremsstrahlung and cyclotron cooling in the accretion column, and the degree of X-ray reflection off the WD surface to precisely represent the dynamic nature of gas in the accretion column.  Our timing analysis characterized broadband X-ray lightcurves well but found no X-ray QPO signal from EF Eri. To test the current accretion column stability models and MHD simulations, searching for X-ray QPOs in the \nicer\ timing data will be more optimal for higher count rates and soft X-ray band coverage.    
    This observation represents the first successful WD mass determination provided by the \nustar\ Polar ToO program and has illustrated the importance of future hard X-ray observations upon other polars.  The high-quality X-ray spectrum of additional polars combined with our sophisticated multi-temperature model will provide valuable insight into the complex nature of the polar accretion column, as well as general relationships between key polar parameters such as WD mass and magnetic field strength. 

\section{Acknowledgments}
Support for this work was provided by NASA through \nustar\ Cycle 9 (NNH22ZDA001N-NUSTAR) and NASA ADAP program (NNH22ZDA001N-ADAP) grants. 
M.T.W. acknowledges support by the NASA Astrophysical Explorers Program. 
VFS thanks the Deutsche Forschungsgemeinschaft (DFG) for financial support (grant WE 1312/59-1). DAHB acknowledges support from the National Research Foundation of South Africa. 
We acknowledge the variable star observations from the AAVSO International Database contributed by observers worldwide and used in this research. We thank Jon Kwong, Chris Deng, and Ciro Salcedo for their contributions to the {\tt MCVSPEC} model, as well as EF Eri timing and spectral analysis. We thank Matteo Bachetti for his assistance with EF Eri timing analysis.  

\bibliography{refs}{}
\bibliographystyle{aasjournal}

\end{document}